\def\draftversion{false}
\RequirePackage{ifthen}
\ifthenelse{\equal{\draftversion}{true}}{
  \documentclass[citeautoscript,floatfix,aps,prx,twocolumn,preprintnumbers,
      superscriptaddress,longbibliography,amsmath,amssymb]{revtex4-1}
      \usepackage{showlabels}
}{
  \documentclass[citeautoscript,floatfix,aps,prx,twocolumn,
      superscriptaddress,longbibliography]{revtex4-1}
      
}

\usepackage{amsmath}
\usepackage{graphicx}
\usepackage{epstopdf}
\usepackage{natbib}
\usepackage{array}
\usepackage{dcolumn}% Align table columns on decimal point
\usepackage{bm}% bold math
\usepackage{siunitx} % ALIGING TABLE COLUMNS BY DECIMAL POINT
\usepackage{subfigure} % Using subfigures

\usepackage{soul}  % \st    for strike-out

\usepackage[usenames,dvipsnames]{color}

\newcommand{\black}{\textcolor{Black}}

\soulregister\cite7

\ifthenelse{\equal{\draftversion}{true}}{
  \marginparwidth 2.7in
  \marginparsep 0.5in
  \newcounter{comm} % counter for commentaries
  \def\commnext{\stepcounter{comm}}
  \def\commtext{{\bf\color{blue}[\arabic{comm}]}}
  \def\commmar{{\bf\color{blue}[\arabic{comm}]}}
  \def\dvm#1{\commnext\marginpar{\small DV\commmar: #1}\commtext}
  \def\cdm#1{\commnext\marginpar{\small CED\commmar: #1}\commtext}
  \def\msm#1{\commnext\marginpar{\small MS\commmar: #1}\commtext}
  \def\asm#1{\commnext\marginpar{\small AS\commmar: #1}\commtext}
  \def\miq#1{\commnext\marginpar{\small MR\commmar: #1}\commtext}
  \def\mlab#1{\marginpar{\small\bf #1}}
  
}{
  \def\dvm#1{}
  \def\cdm#1{}
  \def\msm#1{}
  \def\asm#1{}
  \def\miq#1{}
  \def\mlab#1{}
  
}
\begin{document}

\title{
\black{
Translational covariance of flexoelectricity %in inhomogeneous ferroelectric structures
at ferroelectric domain walls}
%: \\
%Predicting atomic structures via first-principles macroscopic theory
}
% of ferroelectric domain walls: \\
%Flexoelectricity }

\author{Oswaldo Di\'eguez}
\affiliation{Department of Materials Science and Engineering, 
Faculty of Engineering, Tel Aviv University, Tel Aviv 6997801, Israel}
\affiliation{Theory and Simulation Group, Catalan Institute of Nanoscience
and Nanotechnology (ICN2), CSIC, BIST, Campus UAB, Bellaterra, Barcelona
08193, Spain}
%\email{dieguez@tau.ac.il}

\author{Massimiliano Stengel}
\affiliation{Institut de Ci\`encia de Materials de Barcelona 
(ICMAB-CSIC), Campus UAB, 08193 Bellaterra, Spain}
\affiliation{ICREA - Instituci\'o Catalana de Recerca i Estudis Avan\c{c}ats, 08010 Barcelona, Spain}
\email{mstengel@icmab.es}

\date{\today}

\begin{abstract}
\black{
Macroscopic descriptions of ferroelectrics have an obvious appeal 
in terms of efficiency and physical intuition. 
Their predictive power, however, has often been thwarted by the lack of a systematic 
procedure to extract the relevant materials parameters from the microscopics. 
Here we address this limitation by establishing an unambiguous two-way mapping 
between %\st{continuum} 
spatially inhomogeneous fields %(polarization and elastic displacement) 
and discrete lattice modes.
%microscopic lattice modes.
%
%establishing a rigorous mapping between the continuum fields and 
%discrete lattice modes; this naturally yields the continuum model 
%parameters via a long-wavelength expansion of the interatomic
%force constants.
%
%deriving the continuum equations from a rigorous long-wavelength expansion of
%the interatomic force constants, 
%as long-wavelength limit of the interatomic force constants.
%long-wavelength limit of
%the crystal Hamiltonian, which naturally yields the 
%via a pe
%such a procedure yields
This yields a natural treatment of  %\st{macroscopic regime} 
gradient couplings in the macroscopic regime
via a long-wavelength expansion of the crystal Hamiltonian. 
%with the established first-principles theory of flexoelectricity. %treatment of gradient effects.
%
Our analysis reveals an inherent arbitrariness in both the
flexoelectric and polarization gradient coefficients, which we 
ascribe to a translational freedom in the definition of the 
polar distortion pattern.
%
%The converse flexoelectric effect produces a spontaneous 
%Ferroelectric domain walls constitute an attractive
%playground for studying the bulk flexoelectric effect in a context 
%that is free from extrinsic surface contributions.
%
%Here we show, by constructing a macroscopic continuum 
%model of the wall as a well-defined approximation of the 
%first-principles crystal Hamiltonian, that the flexocoupling 
%coefficient is plagued by an inherent arbitrariness, which we 
%rationalize as a weight freedom in the definition of the local 
%strain field.
%
Remarkably, such arbitrariness cancels out in all
physically measurable properties (relaxed atomic structure and 
energetics) derived from the model,
%with an analogous ambiguity 
%in the definition of the local strain field, 
pointing 
%thus yielding unambiguous
%answers for the 
%resulting in a well-defined 
%theory overall.
%
%This result points 
to a \emph{generalized translational covariance} in the continuum
description of inhomogeneous ferroelectric structures.
% the contributions
%of flexoelectricity and correlation are individually ill-defined, but their 
%consistent treatment is crucial for guaranteeing that  are correctly represented.
%
We demonstrate our claims with extensive numerical tests on 180$^\circ$ domain walls 
in common ferroelectric perovskites, finding excellent agreement between the continuum 
model and direct first-principles calculations. % of atomic structures and energies.
}
\end{abstract}

 %Lattice dynamics: first-principles theory
\maketitle

%%%%%%%%%%%%%%%%%%%%%%%%%%%%%%%%%%%%%%%%%%%%%%%%%%%%%%%%%%%%%%%%%%%%%%%%%%%%%%%%

\section{Introduction}

\black{
Spatially inhomogeneous structures in ferroelectrics such as
domain walls, vortices, etc. have been the subject of 
intense research
%received growing
%attention 
in the past few years,~\cite{Seidel-16,Chen21AdvMat} because of their 
emerging physical properties and nontrivial topology. %~\cite{otros}
%
%From the point of view of fundamental research, the central question
%consist in identifying 
Considerable efforts are currently directed at identifying
the physical mechanisms that govern the stability of the observed 
patterns and their response to external probes.
% is currently a subject
%of intense research.
%
In addition to the well-known factors related to the electrical and
mechanical boundary conditions, 
%recent studies have revealed
%the importance of 
%Nonuniform strain fields are ubiquitous in all these systems, which
%means that 
flexoelectricity (describing the coupling between polarization and
strain gradients~\cite{pavlo_review,gustau1,gustau2,cross})
has been receiving increasing attention in this context.
%in this context is becoming increasingly clear.
%
On one hand, the flexoelectric coupling contributes substantially to 
the gradient energy, to the point that a transition to a modulated
phase may occur if sufficiently strong.~\cite{Axe-70,Pottker-16,Tagantsev-13}
%often plays an 
%important role in the stabilization of the observed patterns.~\cite{kalinin-17,Tagantsev-13}
%
%Most importantly, it 
On the other hand, flexoelectricity endows the spatial gradients of the main
order parameters with potentially useful functionalities, e.g., a spontaneous
polarization at ferroelastic twin boundaries,~\cite{Schiaffino-17,salje-16} and 
a spontaneous strain at ferroelectric walls via the converse effect.~\cite{Yudin-13,Wang-20}
}

\black{
In light of these findings, improving our understanding 
of the interplay between flexoelectricity and ferroelectricity
appears as essential for future progress.
%
%A notable step in this direction consists in the predicted relation~\cite{Yudin-13,Wang-20}
%between the flexocoupling coefficient and the spontaneous elastic offset in
%a vicinity of a domain wall.
%In the latter case, in particular, the impact of bulk flexoelectricity 
%can be related to the relaxed atomic positions in the vicinity of 
%the wall,~\cite{Wang-20} which in principle
%This, at first sight, provides a means of extracting the former quantity 
%from an experimental electron microscopy image of the 
%latter~\cite{kalinin-17} via an analysis of the observed
%atomic structure of the wall.~\cite{Wang-20}
%
%For this reason, ferroelectric domain walls appear as a unique
%playground to gauge the theoretical predictions of the bulk
%flexoelectric effect, in a context that is 
%manifestly free from extrinsic 
%surface contributions.~\cite{Tagantsev-12,artgr}
%
%In light of 
Given the advances that the first-principles theory of 
flexoelectricity has made since the pioneering works of Resta~\cite{resta-10} and 
Hong {\em et al.}~\cite{hong-10}, such a goal appears now well within reach.
%an opportunity appears especially 
%appealing.
% in 2010, an impressive 
%progress.~\cite{chapter-15}
%
}
As of early 2020, a complete calculation of the bulk flexoelectric
tensor can be carried out\black{~\cite{Royo2019,Royo-22}} with the latest
release of 
the publicly distributed ABINIT\black{~\cite{abinit,ABINIT2020}} package,
\black{providing 
\emph{in principle} a solid theoretical reference for the interpretation 
of the experimental data.~\cite{kalinin-17,Wang-20}
Unfortunately, these studies have also revealed that the 
bulk flexoelectric coefficients are ill-defined as stand-alone 
material properties. 
More specifically, their definition is plagued by two distinct 
ambiguities, which are respectively related to the treatment
of elastic and electrostatic fields in the 
long-wavelength expansion.~\cite{Hong-13,artlin,Royo-22}
This fundamental limitation prevents a straightforward
%
% complicates
%the 
comparison between theoretical and experimental results, 
as further considerations are needed to make sure that  
the calculated values relate to what is being measured in
a physically meaningful way. %portrait of
%corresponds to 
%what is actually measured.
}

%A clear understanding of such ambiguities is a prerequisite 
%to any attempt at interpreting the experimental data based 
%on the first-principles results, in order to make sure that
%their comparison is physically meaningful.
%are clearly problematic when comparing the calculated 
%values to the experimental measurements, and requires a careful understanding
%of their fundamental origin to make sure that 
%consideration
%of the macroscopic models that are used to interpret the data.

%when incorporating the calculated
%values in a macroscopic model, as one must make sure that the
%physical predictions are independent of the specific conventions
%that are used in either context.

\black{
If they only concerned the specifics of how the flexoelectric effect is
defined and measured, these issues would be of limited importance.
On the contrary, the arbitrariness of the coupling coefficients is 
problematic in a much broader context, as it questions the validity
of the widely popular Landau-Ginzburg-Devonshire (LGD) theories of
ferroelectrics,
%
%The latter, 
a cornerstone of the theoretical understanding of
inhomogeneous polar structures for several decades.
%which have been a cornerstone of the 
%
At a domain wall, standard LGD models predict~\cite{Yudin-13,Wang-20} a dependence of 
both the energy and structure on the flexocoupling coefficient
via the converse effect, which associates a uniform strain with
a gradient of the polar order parameter.
The obvious question is then: how can we trust such physical 
predictions once we know that one of the materials properties 
on which they depend is ill- (or at least nonuniquely) 
defined?
}

\black{
%From the formal point of view, the central issue for achieving this goal consists
%in 
To formulate the problem on firm theoretical grounds, %the first goal
the first challenge consists in establishing
%one needs first of all to establish
%consists in establishing 
a rigorous two-way mapping between microscopic 
degrees of freedom and macroscopic order parameters.
}
In the case of spatially homogeneous crystal phases, 
such a task poses limited conceptual \black{issues}:
Building effective low-energy Hamiltonians in terms of the physically 
relevant lattice distortions (in perovskite crystals these typically 
include polarization, strain and antiferrodistortive oxygen tilts) is 
now common practice~\cite{Zhong-94,Zhong-95,Ghosez-00,Kornev-07} within 
the \emph{ab initio} community.
Whenever these degrees of freedom are no longer constant over space, 
however, \black{many subtleties arise, and the partition of the energy into 
different macroscopic contributions generally becomes nonunique.~\cite{Stengel-16}}
%re are a number of subtleties
%that need to be carefully considered.
%
%These concern the definition of the local strain, of the polar 
%distortion pattern and the flexoelectric tensor itself, all of which 
%imply making arbitrary choices along the way. 
%
%As we shall see, these apparently innocent choices have a
%drastic impact on both the continuum model parameters and the
%domain-wall solution as expressed in terms of field variables.
%
The question, then, is: are there specific criteria for 
ensuring that the result is physically meaningful? And, 
once we have solved the continuum equations, how can we verify
that our solution is consistent with the ``training model'',
i.e., our first-principles engine?

Here we show, by deriving the continuum equations and
parameters \black{via a rigorous long-wavelength approximation 
of the first-principles lattice Hamiltonian, that the above difficulties 
can be traced back to a \emph{translational freedom} 
in the definition of the polar distortion pattern.
%
%In addition to providing an intuitive interpretation of
%the ambiguities that were pointed out before in the context
%of flexoelectricity, 
As a consequence of such freedom not only the 
flexocoupling, $f$, but also the polarization gradient
coefficient, $G$ (entering the continuum functional via the
squared gradient of the polarization field)
is affected by an inherent arbitrariness in its
definition.
% as well defined as 
%stand-alone physical properties.
%
Crucially, we find that the respective ambiguities in $f$ and $G$
%affects both the flexoelectric coefficient,
%consistent with earlier observations,~\cite{Hong-13,artlin,Royo-22}
%the aforementioned ambiguity \black{in the
%definition of the flexoelectric coefficient}
% have a common origin,
%and can be understood as a 
%local center of mass.
%
%
%Remarkably, the ambiguities 
cancel out exactly in any physical prediction of the continuum model, 
implying that a consistent treatment of both terms is essential for
the overall theory to work.}
%thereby generalizing the arguments of Hong and Vanderbilt~\cite{Hong-13} 
%to cases where both the strain \emph{and} the polarization are 
%allowed to be inhomogeneous in space.} % of ferroelectric crystals.
%provided that a 
%unique convention choice is made throughout its construction. %the calculation of all continuum coefficients. 

\black{Of particular note, the aforementioned arbitrariness directly affects the 
definition of the elastic displacement field (ad hence the strain), which we
find to be nonunique.}
% that emerges as a solution of the continuum 
%equations 
%depends on some arbitrary choices of 
%dimensionless weights that one uses to perform a local average 
%of the individual atomic displacements. 
%Such arbitrariness is, however, only apparent:} 
\black{Our long-wave approach to continuum theory, however, yields unique 
answers for the domain-wall structure once the local field amplitudes 
are converted back into atomic distortions, enabling a straightforward
validation of the method against direct density-functional theory (DFT) calculations.}
\black{We %demonstrate our claims 
illustrate this point} by calculating 180$^\circ$ ferroelectric
walls in six different perovskite materials, finding answers 
that are within 10--20\% of the ``exact'' result. Given
the extreme (one-cell thick) abruptness of the structures, 
we regard this as a severe test for a continuum approach,
and such an accuracy exceptionally good.
By calculating domain walls under hydrostatic pressure in BaTiO$_3$
we also demonstrate the exactness of our theory in the limit
of smooth domain walls.

%relaxed atomic structure
%and energies of a domain wall, 
%which can be directly validated
%with the results of 
%
%Thus,  the results for the energetics and structure are
%unique 
%physical answer provided by our approach are robust and consistent 
%once the continuum fields are converted back into 
%
From our results, a new paradigm emerges in the construction 
of continuum models of ferroics: the invariance of the Landau-Ginzburg-Devonshire 
free energy with respect to a number of \emph{generalized gauge
transformations} of the parameters and fields.
This implies abandoning the widespread belief that such 
parameters \black{(e.g., the flexoelectric coefficient) and fields (e.g., 
the local strain) be well-defined physical properties of the crystal.}
It also emphasizes the need for an intimate connection between microscopics 
and macroscopics in order to achieve a qualitatively sound picture.

This work is organized as follows. 
In Section~\ref{sec:theory} we address the theoretical issues that arise in the design
of continuum models, focusing on the aforementioned ambiguities 
in the definition of local strains, polar distortion patterns, and
flexoelectric coefficients.
In Section~\ref{sec:results} 
we present our numerical tests on
180$^\circ$ domain walls in perovskite oxides, alongside with a 
detailed validation against the results of direct first-principles 
calculations. 
In Section~\ref{sec:discussion} we discuss the implications of
our findings in the context of the relevant literature.
We summarize our work and present our conclusions in Section~\ref{sec:conclusions}.

%%%%%%%%%%%%%%%%%%%%%%%%%%%%%%%%%%%%%%%%%%%%%%%%%%%%%%%%%%%%%%%%%%%%%%%%%%%%%%%%

\section{Theory}

\label{sec:theory}

\begin{figure}
\centering
\includegraphics[width=80mm,angle=0]{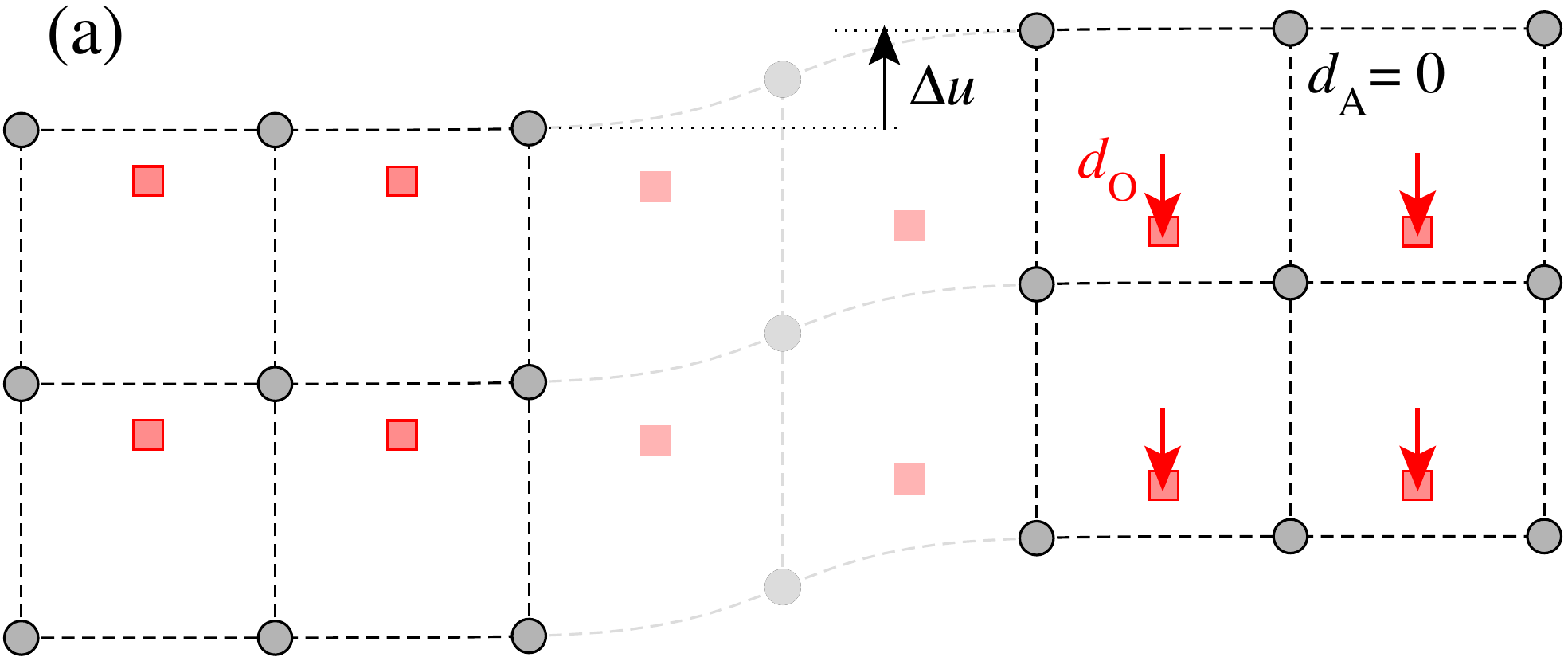} \\
\vspace{10mm}
\includegraphics[width=80mm,angle=0]{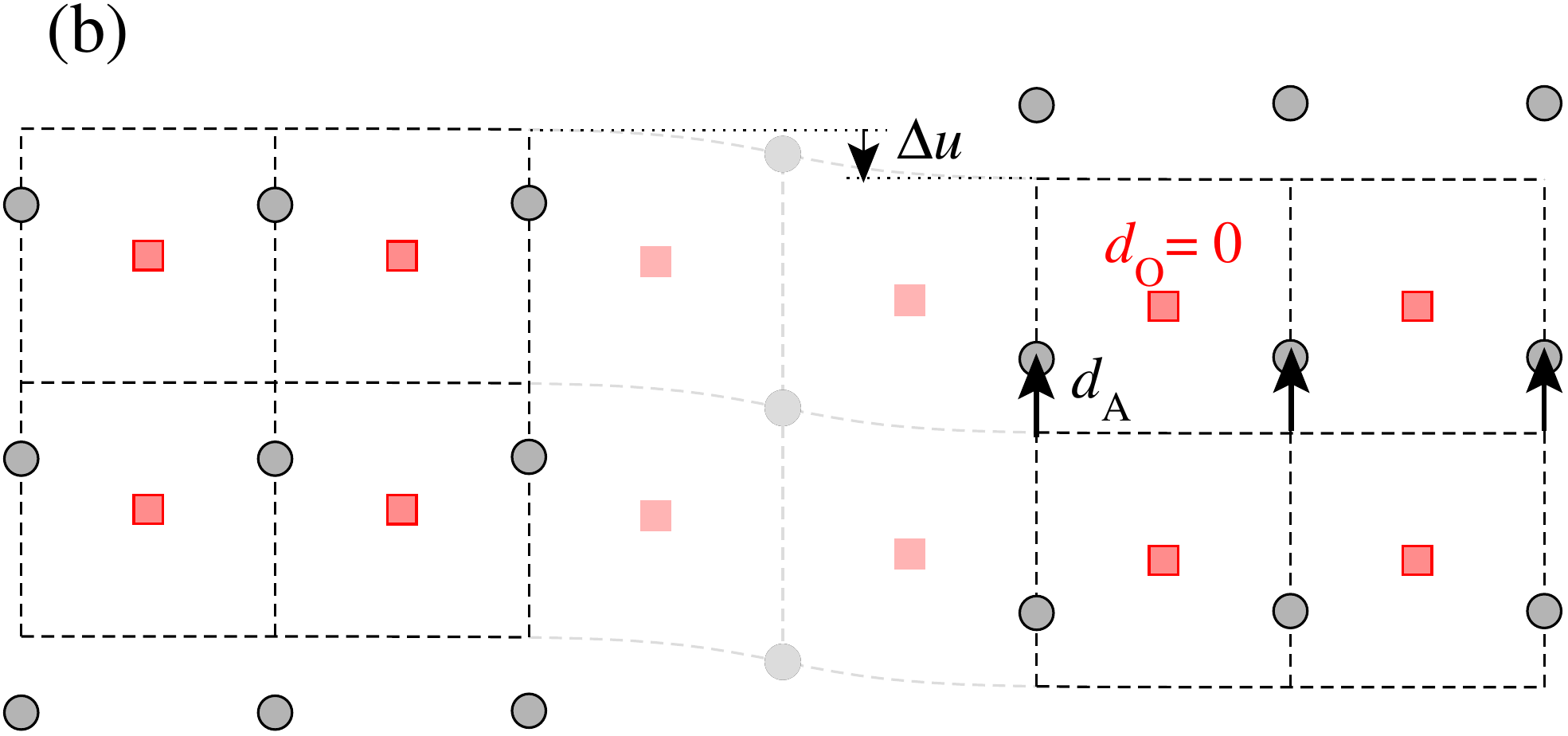} 
\caption{(Color online.) Schematic model of a [100]-oriented 
ferroelectric domain wall in a perovskite crystal. For simplicity, only atoms in AO 
planes \black{perpendicular} to the wall are shown as red circles (A) and black squares (O), respectively.
The elastic displacement of the crystal cells (dashed lines) is also shown.
The ferroelectric distortion of the lattice deep within the domain is indicated
by red and black arrows, respectively for the displacements ($d_\kappa$) of the A and O sublattice
with respect to their high-symmetry positions. The macroscopic shift of the crystal cell, $\Delta u$,
is also shown. Displacements are exaggerated for illustration purposes. \black{(a) and (b) illustrate 
how two different choices for $d_\kappa$ and $\Delta u$ can be made for the same atomic structure.}}
\label{cartoon}
\end{figure}

\subsection{Statement of the problem}

Consider a 180$^\circ$ domain wall in a ferroelectric crystal 
as schematically illustrated in Fig.~\ref{cartoon}. 
The outer extremes of both panels correspond to the oppositely
oriented ferroelectric domains, where the polarization ($P$) saturates to 
its bulk value; in the central domain-wall region $P$ %the polar distortion 
transitions from negative to positive values when moving from left 
to right.
The atomic structure far from the wall is well described 
in terms of a periodically repeated crystalline cell. 
Such a structure can be readily obtained from a bulk calculation:
one typically starts from the reference configuration, breaks
the centrosymmetry by hand (e.g. by displacing atom A upwards or
atom O downwards by a small amount), and lets the atoms relax to 
their polar ground state.
Note that one can perform the relaxation while fixing either A or O  
to their original locations; while the resulting distortions, 
$d_{\rm A,O}$ differ [compare panels (a) and (b) of Fig.~\ref{cartoon}],
the two structures are related by a rigid translation of the whole lattice,
and are therefore degenerate. % at the level of the periodic bulk domain.

Once the domain walls forms, the two oppositely oriented domains
no longer enjoy translational invariance separately: the wall
lifts the indeterminacy modulo a rigid shift of the cell, and 
uniquely sets the registry between the two oppositely polarized
half-lattices.
The shift that we must add to the calculated bulk 
atomic structure to correctly align the two semi-infinite regions
can conveniently be rationalized in terms of the \emph{elastic displacement 
field}, $u(x)$, which undergoes a jump, $\Delta u$, when moving across the 
wall along the normal direction ($x$).
Thus, the spontaneous alignment between the two domains can be 
understood physically as an electromechanical effect, where the
elastic degrees of freedom emerge as secondary consequence of the 
primary polar distortion of the lattice.
And indeed, recent works~~\cite{Yudin-13,Wang-20} have clarified that 
the net elastic displacement is due to \emph{converse flexoelectricity}, 
and $\Delta u$ can be related linearly to the flexocoupling coefficient(s) of the 
crystal in its cubic reference phase.

Earlier works, however, have overlooked the central conceptual
issue with the above interpretation. Since the ferroelectric distortion
pattern within the bulk domains (quantified here by $d_\kappa$, with 
$\kappa$=A,O) is ill defined, then the amount of elastic 
displacement, $\Delta u$, that we must incorporate to 
obtain the correct registry between the left and right half-lattices
is also ill defined.
This is obvious by looking at Fig.~\ref{cartoon}(a--b), 
where we compare two different choices for $d_\kappa$ and
$\Delta u$. Clearly, the total atomic distortions within each
domain, given by $\pm(d_\kappa + \Delta u/2)$, are the same
in (a) and (b).
And yet, what we understand as the ``macroscopic elastic offset'' 
between the domains, $\Delta u$, markedly differs.
This situation is paradoxical in light of the widespread 
assumption that the local elastic displacement field
(and hence the local strain) be a physically unambiguous degree
of freedom of the crystal.
%
%The flexoelectric coefficients, which are ultimately 
%responsible for the appearance of the
%elastic offset at the wall~\cite{Yudin-13,Wang-20})
%are commonly regarded as well-defined crystal properties, too. 
%
\black{
The sketch of Fig.~\ref{cartoon}
seems to disprove such an interpretation.
As we have anticipated in the introduction,
abandoning such long-established paradigm brings
about a number of conceptual troubles; we 
shall address them in the remainder of this 
work.
%: how can we trust, for 
%example, macroscopic theory while knowing that its very foundations
%are plagued by ambiguities? Is there a criterion we can use to make
%sure that the physical answers that we extract from such a theory
%are sound? And, ultimately, is there any use at studying  
%flexoelectricity once we know that its contributions to the physical
%properties of materials are ill-defined?
%
%We shall address the above questions in the remainder of
%this work.
}

\subsection{\black{Macroscopic theory}}

\label{sec:landau}

%Consider a 180$^\circ$ domain wall in a proper ferroelectric. 
%We shall consider  domain-wall model 
%The 
%polarization is oriented along $y$, and is parallel to the domain
%wall plane. The  
%
%We shall also consider the , 
%\begin{equation}
%\varepsilon = \frac{\partial u}{\partial x}.
%\end{equation}
%
%\black{In this Section we shall review the established theory~\cite{yudin-13} of a
%
\black{To frame our discussion, in this Section we recap the established~\cite{Yudin-13} macroscopic 
theory for an Ising-like 180$^\circ$ ferroelectric domain-wall as illustrated in Fig.~\ref{cartoon}.
For simplicity, we shall exclusively focus on the transverse ($y$) component of the polarization:
%since
%this is justified 
%, where 
Longitudinal ($x$) components are  
%strongly suppressed by depolarizing effects 
typically small in perovskite ferroelectrics,
%
%(This removes the need for 
and would require an explicit treatment of the electrostatic energy,
which is nontrivial in the flexoelectric case.~\cite{Stengel-16}
%components are 
We also restrict our attention to a single strain component, the $xy$ shear,
which is the most relevant one to our present scopes. (This implies neglecting 
the tetragonal distortion of the crystal cells deep within the domains.)
%
%which means that the only active strain component is the $xy$ shear.
Within these assumptions, the simplest free-energy functional to describe the 
problem is}
%treat this problem 
\begin{equation}
\label{free}
F(P,\varepsilon) = \frac{C}{2} \varepsilon^2 + \frac{A}{2} P^2 + \frac{B}{4} P^4 + f \varepsilon \frac{\partial P}{\partial x}
 + \frac{G}{2} \left( \frac{\partial P}{\partial x} \right)^2.
\end{equation}
\black{Here $P$ is the parallel ($y$) component of the polarization, while the normal to the wall is 
indicated as $x$;
\begin{equation}
\varepsilon = \frac{\partial u}{\partial x}
\end{equation}
is the shear ($xy$) component of the strain, defined as the
$x$-derivative of the parallel ($y$) component of the displacement field, ${u}$;}
$C$ is the elastic constant ($C_{44}$ component), $A$ and $B$ are the standard parameters
of the \black{homogeneous} Landau potential, $f$ is the flexocoupling coefficient, and $G$ is the \black{polarization 
gradient} coefficient.
Note that the flexoelectric coupling is written here in terms of the converse effect
(uniform strain in response to a $P$-gradient); it is related to the direct effect
and to the standard Lifshitz-invariant formula via simple integrations by parts,
% What is the exact meaning of the arrows? That the values of f are different
% for each expression? If not, maybe it would be more clear to write that when
% only the bulk is considered, the three terms are equal, and use the usual
% symbol "="?
\begin{equation}
f \varepsilon \frac{\partial P}{\partial x} \,\, \simeq \,\,
-f \frac{\partial \varepsilon }{\partial x} P \,\, \simeq \,\,
\frac{f}{2} \left( \varepsilon \frac{\partial P}{\partial x}  - \frac{\partial \varepsilon }{\partial x} P \right).
\end{equation}
(The difference between the three expressions consists in surface terms, irrelevant for
the present bulk theory.)

By imposing the stationary condition %It is clear that the strain can be integrated out, 
\begin{equation}
\frac{\partial F}{\partial \varepsilon} = 0,
\end{equation}
the strain can be integrated out, which immediately leads to 
%This condition immediately leads to 
the following result,~\cite{Yudin-13} %for the strain and the displacement field,
% If the arrow means "by direct integration over x", I would write that,
% creating two equations.
\begin{equation}
\varepsilon = -\frac{f}{C} \frac{\partial P}{\partial x} \,\, \to \,\, u = -\frac{f}{C} P.
\label{eq_u}
\end{equation}
(We have eliminated a trivial integration constant in $u$ by imposing that 
$u$ vanishes whenever $P=0$.)
Therefore, the displacement field at a ferroelectric domain wall adopts the 
exact same spatial profile as the polarization, except for the $-f/C$ 
scaling factor. 
%
%One can also extract
From Eq.~(\ref{eq_u}) one then can extract the net elastic offset, 
$\Delta u$, that we have introduced in the previous Section,
\begin{equation}
\label{offset}
\Delta u = -2 \frac{P_0 f}{C}.
\end{equation}

After eliminating the strain, we obtain the following simpler expression for
the free energy,
\begin{equation}
\label{renorm}
F(P) = \frac{A}{2} P^2 + \frac{B}{4} P^4 
 + \frac{\tilde{G}}{2} \left( \frac{\partial P}{\partial x} \right)^2, \qquad \tilde{G} = G - \frac{f^2}{C}.
\end{equation}
The condition for stability is that the renormalized \black{polarization gradient} coefficient be positive, $GC > f^2$.
(This criterion is well known: if $f$ is large enough, the system becomes unstable
and an incommensurate transition to a modulated state may occur.~\cite{Pottker-16,Axe-70,Tagantsev-13})
The equation of state is given by the stationary condition with respect to $P$,
\begin{equation}
AP + BP^3 - \tilde{G} \frac{\partial^2 P}{\partial x^2} = 0.
\end{equation}
%
%(Second derivative wrt $x$ is indicated as a double prime superscript.)
We shall attempt a trial solution of the type
\begin{equation}
\label{eqtanh}
P(x) = P_0 \tanh \left( \frac{x}{\xi} \right).
\end{equation}
After a few steps of straightforward algebra, we arrive at
\begin{equation}
\label{eqxi}
P_0^2 = -\frac{A}{B}, \qquad \xi^2 = \frac{2 \tilde{G}}{|A|}.
\end{equation}
$P_0$ is determined by the bulk Landau potential, while
$\xi$ is a length, and has the obvious physical meaning of domain wall thickness.

%We shall need the derivatives of the hyperbolic tangent,
%\begin{equation}
%\frac{d}{dx} \tanh(x) = 1-\tanh^2(x), \qquad \frac{d^2}{dx^2} \tanh(x) = -2 \tanh(x) [1-\tanh^2(x)].
%\end{equation}
%
%So, the equation of state reduces to [$t=\tanh(x/\xi)$]
%\begin{equation}
%A P_0 t + B P_0^3 t^3 +  \frac{2 \tilde{G}}{\xi^2} P_0 t (1-t^2) = 0.
%\end{equation}
%%
%Dividing by $P_0 t$, and inserting the value of $P_0^2$, we 
%have
%\begin{equation}
%A - A t^2 +  \frac{2 \tilde{G}}{\xi^2} (1-t^2) = 0.
%\end{equation}
%
%The condition for the solution is then
%\begin{equation}
%\frac{2 \tilde{G}}{\xi^2} = -A, \qquad \xi^2 = \frac{2 \tilde{G}}{|A|}.
%\end{equation}
%

%\section{Energy}

The domain-wall energy per unit area can be obtained by integrating the free energy 
density along the normal to the wall,
\begin{equation}
W = \int dx \, [F(P(x)) - F_0],
\end{equation}
where $F_0$ is the energy density of the monodomain ground state.
%
% By using the solution of the previous Section, one arrives at 
One arrives at 
%
%
%can write the integrand as
%\begin{equation}
%\Delta F(P) = \frac{A}{2} (P^2-P_0^2) + \frac{B}{4} (P^4-P_0^4)
% + \frac{\tilde{G}}{2} \left( \frac{\partial P}{\partial x} \right)^2,
%\end{equation}
%which leads to
%\begin{equation}
%\begin{split} 
%\Delta F (P) &= -\frac{A^2}{2B} (t^2 - 1) + \frac{A^2}{4B}(t^4 - 1) + \frac{A^2}{4B}(t^2 - 1)^2 \\
%             &= \frac{A^2}{4B} (t^2 - 1) [ -2 + (t^2 + 1) + (t^2 - 1) ] \\
%             &= \frac{A^2}{2B} (t^2 - 1)^2.
%\end{split}
%\end{equation}
%
%The integral simplifies, therefore, to
\begin{equation}
\label{eqdwe}
W = \frac{A^2}{2B} \xi \int_{-\infty}^{+\infty} dx \, {\rm sech}^4 (x) = \frac{A^2}{2B} \xi \frac{4}{3} = \frac{8 }{3} |F_0| \xi,
\end{equation}
where $F_0=-A^2/4B$ is the bulk energy density.
[Since $F_0$ is an energy per unit volume, Eq.~(\ref{eqdwe}) correctly describes $W$ 
in units of energy per length squared.]
%
%This means that the domain wall energy corresponds to
%a thickness of about $2\xi$ where the polarization vanishes.
%
%
%Closer inspection of the above formulas reveals that $W$ %the domain wall energy
%is equally distributed between two separate contributions: (i) the bulk on-site 
%energy, coming from the suppression of the polarization in the domain wall region;
%(ii) the gradient energy, consisting in the cost of the inhomogeneity in the 
%polarization profile. 
%
%We have, in particular,
%\begin{equation}
%\label{eqw}
%W=W_1 + W_2, \qquad W_{1,2} = \frac{A^2}{4B} \xi \frac{4}{3} = -\frac{4}{3} \xi F_0.
%\end{equation}
%
%From the point of view of the physical dimensions, $F_0$ is an energy per unit volume;
%thus, since $\xi$ is a length, $W$ is an energy per unit area (as it should be).

An interesting consequence of the above derivations is that %the relevant 
\black{the main} physical \black{properties of the wall}  %observables
%i.e., the domain wall 
(the thickness, $\xi$, and the energy, $W$), depend on the inhomogeneous coefficients 
only via the renormalized gradient coupling, $\tilde{G}$, \black{while other features
(the elastic offset, $\Delta u$) explicitly depend on the flexocoupling, $f$.}
This means that we can replace $f$ and $G$ with arbitrary numbers, provided that $G-f^2/C$ 
retains the original value, and extract the same physical answers for $\xi$ and $W$; %from our theory 
\black{$\Delta u$, on the other hand, is not invariant with respect to such a transformation.}
This property of Eq.~(\ref{free}) may appear at first sight as a mathematical curiosity, 
but has profound physical implications in relation to the paradox illustrated in 
Fig.~\ref{cartoon}; we shall explore them in the following subsections.

\subsection{Direct mapping to the microscopics}

\label{sec:direct}

\black{To test the validity of Eq.~(\ref{free}) in real systems,
%In order to discuss the above results in our specific context,
%it is important to 
we need %first of all 
to establish a microscopic interpretation of the 
order parameters entering the continuum functional.
%continuum fields ${\bf u}({\bf r})$ and ${\bf P}({\bf r})$.
%: without a clear mapping 
%between continuum and atomistics we would be unable 
%
In full generality, we use the following two-step
procedure.}
%consider a set of continuum fields
%$v_{\nu}({\bf r})$, where 
%
%
First, we express the individual atomic displacements as continuum functions
of the real-space coordinates, ${\bf r}$, \black{via a linear transformation of the
relevant vector fields, % in the system, 
$v_{\nu \beta}({\bf r})$,}
%by mapping the relevant 
%order parameters
%onto a basis of individual atomic displacemen indexed by $\kappa$,
\begin{equation}
\label{basis}
\black{u_{\kappa \alpha}({\bf r}) = \sum_{\nu \beta} v_{\nu \beta}( {\bf r})T_{\nu\beta,\kappa\alpha},
 %  \sum_\beta u_\beta( {\bf r}) u^{(\beta)}_{\kappa\alpha}  
 %   +  \sum_\beta P_\beta( {\bf r}) \, p^{(\beta)}_{\kappa\alpha}.
    } 
\end{equation}
%
%Here %$\nu=u,P$ is either
%the elastic displacement or the ferroelectric polarization; 
%$\beta$ runs over their respective Cartesian components;
%%
%$v^{(\beta)}_{\kappa\alpha}$ 
\black{($T_{\nu\beta,\kappa\alpha}$ is the transformation matrix of the mapping,  %$\langle \kappa \alpha | v_\beta \rangle$ 
describing the displacement of the sublattice $\kappa$
along $\alpha$ for a \black{unit amplitude} %$\beta$-component 
of ${v}_{\nu \beta}$, where $\beta$ runs over the Cartesian components 
of the vector field ${\bf v}_\nu$.)}
Second, we write the displacement of the atom in the $l$-th cell by 
sampling the atomic displacement fields at the 
undistorted lattice sites, ${\bf R}_{l\kappa}^{(0)}$,
% In order to make things more clear, I would suggest giving some examples
% where the first term of the rhs of Eq 13 is not zero, and where the
% braket of the second term is not proportional to Kronecker's delta.
\begin{equation}
\label{micro}
u^l_{\kappa \alpha} = u_{ \kappa\alpha}( {\bf R}_{l\kappa}^{(0)}).
\end{equation}
\black{The combination of Eq.~(\ref{basis}) and Eq.~(\ref{micro}) endows the
order parameters with the meaning of envelope functions, generally
smooth on the scale of the interatomic spacings, that modulate 
a cell-periodic displacement pattern of the atoms over the 
volume of the crystal.}
%This means that the atom $l\kappa$ displaces according
%to the amplitude of the continuum field, ${\bf u}$ or ${\bf P}$, 
%at the ${\bf R}_{l\kappa}^{(0)}$ point, times a %sublattice-dependent

\black{We assume a perovskite-structure lattice henceforth and
focus on two specific order parameters, $\nu=u,P$, corresponding to 
the elastic displacement and ferroelectric polarization.
We identify their respective blocks of the
transformation matrix, ${\bf T}$, 
%corresponding 
%atomic displacement patterns 
with the threefold 
degenerate acoustic and ``soft'' transverse modes of the 
undistorted cubic structure at the zone center.
%
%Given the $T_{1u}$ symmetry of both $u$ and $P$, and a
Within the standard choice of the
coordinate axes, both blocks are diagonal on the Cartesian indices,
%we have %both blocks are diagonal on the Cartesian indices,
%the relevant blocks of the transformation 
%matrix  
\begin{equation}
\label{pb}
T_{u\beta,\kappa\alpha}= u^{(\beta)}_\kappa \delta_{\alpha \beta}, \qquad 
T_{P\beta,\kappa\alpha}= p^{(\beta)}_\kappa \delta_{\alpha \beta}.
\end{equation}
%
%and can be expressed in terms of the 
%where 
$u^{(\beta)}_\kappa$ and $p^{(\beta)}_\kappa$ are sets of three five-dimensional \emph{basis
vectors}, each forming a $T_{1u}$ irreducible representation of the $Pm\bar{3}m$ point group.
Regardless of the microscopics, $u^{(\beta)}_\kappa$ neither depends 
on the sublattice index $\kappa$ nor on the Cartesian index $\beta$, 
as it describes a rigid shift of the cell that is collinear with 
$u_\beta$,~\cite{Stengel-16}
\begin{equation}
\label{ub}
%\langle \kappa \alpha| u_\beta \rangle 
%u^{(\beta)}_{\kappa\alpha} 
u^{(\beta)}_\kappa = 1.
\end{equation}
%
%On the other hand, if $\alpha\neq\beta$, $p^{(\beta)}_\kappa$ and $p^{(\alpha)}_\kappa$ have 
%the same independent entries but differ 
%provide complete information about ${\bf T}$ in our case.}
%
%Since both modes have  modes of the 
%
%(Here the scalar products between bras and kets stand for combined sums over
%an internal sublattice and Cartesian displacement index.~\cite{Zabalo-21})
%
%For example, by using the definition 
%of the elastic displacement vector at the zone center,
%
%
%Regarding the contribution of 
%Regarding the ferroelectric degree of freedom, 
%$\alpha$ and $\beta$ coincide as well (we assume a standard choice of the Cartesian axes) 
%in the case of the ferroelectric polarization as well, 
%and the corresponding block of the
%transformation matrix contains five independent entries,
%
%The distortion vector $p^{(\beta)}_{\kappa\alpha}$, %$\langle \kappa \alpha | P_\beta \rangle$
%on the other hand, refers to a zone-center polar mode of the
%undistorted cubic structure.
% crystals
%the ferroelectric distortion of the lattice.
%on the other hand, contains a factor
%$\langle \kappa \alpha | P_\beta \rangle$, describing %which is defined as 
%the displacement 
%of sublattice $\kappa$ along $\alpha$ that is associated to the $\beta$-component 
%of the macroscopic (uniform) ferroelectric distortion. 
%
%
%and $p^{(\beta)}_{\kappa\alpha}$ 
%describing the displacement of each atom along a given pseudocubic direction,
%\begin{equation}
%\label{pb}
%T_{P\beta,\kappa\alpha}= \delta_{\alpha \beta} p^{(\beta)}_\kappa.
%\end{equation}
%
Eq.~(\ref{basis}) reduces then to the following simplified expression,}
\begin{equation}
\label{basis2}
\black{u_{\kappa \alpha}({\bf r}) = u_\alpha( {\bf r}) + P_\alpha( {\bf r}) p^{(\alpha)}_\kappa
%\sum_{v\beta} v_\beta( {\bf r})v^{(\beta)}_{\kappa\alpha},
 %  \sum_\beta u_\beta( {\bf r}) u^{(\beta)}_{\kappa\alpha}  
 %   +  \sum_\beta P_\beta( {\bf r}) \, p^{(\beta)}_{\kappa\alpha}.
    } 
\end{equation}
%
%At this point, 
where the only remaining task consists in specifying $p^{(\alpha)}_\kappa$.

As detailed in Sec.~\ref{sec:converse}, we require that the homogeneous solution of the 
Landau potential, Eq.~(\ref{free}), reproduce the spontaneous bulk ferroelectric 
distortion pattern, $d^{(\beta)}_\kappa$, via Eq.~(\ref{basis2}):
%
%This condition leads to
\begin{equation}
\label{dbeta}
d^{(\beta)}_\kappa = P_0 p^{(\beta)}_\kappa.
\end{equation}
%
%coincide with the bulk ferroelectric distortion pattern, which we
%use as an operational definition of $p^{(y)}_{\kappa}$.
This condition implies that $p^{(\beta)}_\kappa$
has four independent entries and depends 
on $\beta$ by a permutation of the oxygen indices.~\cite{Hong-13}
Eq.~(\ref{dbeta}) does not lead to a unique solution for
$p^{(\beta)}_{\kappa}$, though: (i)
there is a (trivial) freedom in the choice of the unit in which both ${\bf P}({\bf r})$ and
$p^{(\beta)}_{\kappa}$ are expressed; (ii), because of the translational 
invariance that we have mentioned earlier, $d^{(\beta)}_{\kappa}$ is only defined 
modulo an arbirary shift of the whole lattice.
%along $\beta$ (see Fig.~\ref{cartoon});
%we shall come back to this central point shortly.}
\black{In the following, we shall assume that some choice has been made for (i--ii)
and proceed to deriving all the coefficients entering Eq.~(\ref{free}) in terms 
of microscopic quantities; later on, we shall discuss the implications of (ii)
in regards to the apparent paradox of Fig.~\ref{cartoon}.}

\subsection{Calculation of the coupling coefficients}

The homogeneous coefficients $A$ and $B$ are easy to extract from a 
first-principles calculation: they are readily given by a quartic 
fit of the energy of the primitive cell as a function of the distortion
amplitude along the direction (in configuration space) spanned by 
$p^{(\beta)}_{\kappa}$.
% I think this quartic fit is not true any more: they are computed from 
% two points corresponding to tetragonal and cubic phases.
%$| P_\beta \rangle$.
%
The gradient terms (especially $f$ and $G$) are 
technically more challenging to calculate, in that they 
are defined in terms of spatially modulated (and hence non cell-periodic) 
atomic distortion patterns.
Recent developments~\cite{artlin,Stengel-16,Royo2019} in density-functional perturbation 
theory have overcome these difficulties by applying the long-wavelength method 
to the phonon problem. \black{We shall show in the following that Eq.~(\ref{basis}) %and~(\ref{micro})
directly connects to the formalism of Refs.~\onlinecite{artlin,Stengel-16},
%briefly recap the main results, 
%and present a microscopic expression 
and hence lead to a physically sound definition of $C$, $f$ and $G$.}

%\subsection{Gradient-mediated couplings}

Since all gradient terms are harmonic, \black{we consider a linear-response regime in the 
field amplitude with respect to the high-symmetry cubic phase.
To capture the spatial modulation, it is convenient to work in Fourier space and express
the relevant perturbations of the continuum fields as a constant times a complex phase, e.g., 
${\bf P}({\bf r}) = {\bf P}^{\bf q} e^{i{\bf q}\cdot {\bf r}}$.
Via Eq.~(\ref{micro}), the corresponding lattice distortions can be written   % and~(\ref{micro}),
as linear combinations of monochromatic displacement patterns of the atoms,
\begin{equation}
u^l_{\kappa \alpha} = u^{\bf q}_{\kappa \alpha} e^{i{\bf q}\cdot {\bf R}_{l\kappa}^{(0)}}.
\end{equation}
%
%our starting point are the interatomic force constants % matrix
%calculated in the high-symmetry undistorted geometry,
%\begin{equation}
%\Phi^{l}_{\kappa \alpha, \kappa' \beta} = \frac{ \partial^2 E}{\partial u^0_{\kappa \alpha} \partial u^l_{\kappa' \beta} }.
%\end{equation}
%
%These are defined as second derivatives of the total energy with respect to two atomic 
%displacements, respectively of the atom $0\kappa$ along $\alpha$ and $l\kappa'$ along $\beta$.
%
%Following earlier works, 
The second derivatives of the energy with 
respect to $u^{\bf q}_{\kappa \alpha}$ define~\cite{artlin} 
the \emph{force-constants matrix} at the specified wave vector, {\bf q},
%shall introduce a spatial modulation via the Fourier transform of $\Phi^{l}_{\kappa \alpha, \kappa' \beta}$, 
%which we define as
% Definitions of the \tau quantities as atomic positions are needed.
\begin{equation}
\label{phiq}
\Phi^{\bf q}_{\kappa \alpha, \kappa' \beta} = \frac{\partial^2 E}{\partial u_{\kappa\alpha}^{\bf -q} \partial u_{\kappa'\beta}^{\bf q}},
%\sum_l \Phi^{l}_{\kappa \alpha, \kappa' \beta} e^{i{\bf q}\cdot ({\bf R}_l + \bm{\tau}_{\kappa'} - \bm{\tau}_\kappa)}.
\end{equation}
which provides the formal link to the established density-functional
perturbation theory framework.~\cite{artlin}}

\black{The macroscopic limit of Eq.~(\ref{phiq}) is taken via}
a long-wave expansion~\cite{artlin,Royo-22} in powers of ${\bf q}$,
\begin{equation}
\label{lw}
\Phi^{\bf q}_{\kappa \alpha, \kappa' \beta} = \Phi^{(0)}_{\kappa \alpha, \kappa' \beta} - iq_\gamma
 \Phi^{(1,\gamma)}_{\kappa \alpha, \kappa' \beta} - \frac{q_\gamma q_\delta}{2} \Phi^{(2,\gamma\delta)}_{\kappa \alpha, \kappa' \beta} + \cdots.
\end{equation} 
The zero-th order term is the usual zone-center force-constants matrix in short-circuit electrical 
boundary conditions. \black{(It may be used to compute the homogeneous quadratic coefficient, $A$.)}
The first-order term vanishes in the cubic perovskite reference structure. \black{Finally, the
second-order term allows one to extract the sought-after information about flexoelectricity, 
polarization gradient and elasticity via a projection %of $\Phi^{(2,)}$ 
onto the elastic and polar displacement patterns.}
%
%In particular, 
%we readily obtain the coefficients $C$, $G$ and $f$ in Eq.~(\ref{free}) 
%as 
%
%By using 
\black{In particular, in our specific context of the 
[100]-oriented wall with the polarization oriented along [010],}
Eqs.~(\ref{ub}) and~(\ref{pb}) lead to the following explicit formulas,
% Are the \Phi^(i) the i-th moments of \Phi?
%which immediately leads to the following definitions of the gradient-mediated couplings in Eq.~(\ref{free}),
\black{\begin{subequations}
\label{cfg}
\begin{align}
C =& %-\frac{1}{2\Omega} \langle u_y | \Phi^{(2,xx)} | u_y \rangle = 
     -\frac{1}{2\Omega} \sum_{\kappa \kappa'} \Phi^{(2,xx)}_{\kappa y, \kappa' y}, \\
f =& %-\frac{1}{2\Omega} \langle u_y | \Phi^{(2,xx)} | P_y \rangle= 
     -\frac{1}{2\Omega} \sum_{\kappa \kappa'} \Phi^{(2,xx)}_{\kappa y, \kappa' y} p^{(y)}_{\kappa'}, \\ 
G =& %-\frac{1}{2\Omega} \langle P_y | \Phi^{(2,xx)} | P_y \rangle =
    -\frac{1}{2\Omega} \sum_{\kappa \kappa'} \Phi^{(2,xx)}_{\kappa y, \kappa' y} p^{(y)}_{\kappa} p^{(y)}_{\kappa'},
\end{align}
\end{subequations}}
where $\Omega$ is the volume of the undistorted primitive cell.

\black{To connect with the existing first-principles theory of flexoelectricity, 
%[We assume a (100)-oriented wall, with the polarization oriented along (010)]
%
 %for the elastic \black{and flexocoupling coefficients,
%\begin{subequations}
%\label{sqb}
%\begin{align}
%C        &= -\frac{1}{2\Omega} \sum_{\kappa \kappa'} \Phi^{(2,xx)}_{\kappa y, \kappa' y}, \\
%f        &= -\frac{1}{2\Omega} \sum_{\kappa \kappa'} \Phi^{(2,xx)}_{\kappa y, \kappa' y} p_{\kappa'}, \\
%G        &= -\frac{1}{2\Omega} \sum_{\kappa \kappa'} \Phi^{(2,xx)}_{\kappa y, \kappa' y} p_{\kappa} p_{\kappa'}.
%\end{align}
%\end{subequations}
%One can also 
it is useful to recall the definition~\cite{artlin,Royo-22} of the 
\emph{force-response coefficient}, %$[yy,xx]^{\kappa'}$
\begin{equation}
\label{flexof}
 f^{\kappa'} = -\frac{1}{2} \sum_{\kappa} \Phi^{(2,xx)}_{\kappa y, \kappa' y}
\end{equation} 
as the force on the sublattice $\kappa'$ produced by a gradient of the shear strain.
A comparison between Eq.~(\ref{cfg}) and Eq.~(\ref{flexof}) 
shows that the flexocoupling coefficient is properly 
defined here as the geometrical force on the polar mode produced
by a strain gradient, $f = \sum_{\kappa'} f^{\kappa'} p^{(y)}_{\kappa'} / \Omega$,
consistent with earlier works.~\cite{Stengel-16,Zabalo-21}
Also, the definition of the elastic constant $C$ is consistent with 
the classic result of Born and Huang~\cite{born/huang}, as revisited recently in
a modern electronic-structure context.~\cite{artlin,Royo-22}
Note that $f^\kappa$ and $C$ comply with the ``elastic sum rule'',~\cite{artlin,Royo-22}
\begin{equation}
C  = \frac{1}{\Omega} \sum_{\kappa'} f^{\kappa'},
\end{equation}
as can be easily verified via Eq.~(\ref{cfg}).}

The fact that we obtain the elastic tensor, which is usually regarded as a 
homogeneous coupling parameter, via a similar procedure as spatial dispersion 
coefficients such as $G$ and $f$, might come as a surprise to the reader.
%It might come as a surprise to the reader that we have included the 
%elastic tensor together with $G$ and $f$ in the list of dispersion 
%properties. 
%
This is justified by the fact that a uniform strain is a \emph{gradient} 
of the elastic displacement field, and therefore it formally enters the 
long-wave expansion of the dynamical matrix on the same footing
as a gradient of the polar mode: elasticity, flexoelectricity 
and \black{polarization gradient coupling} all occur at second order 
in the wavevector ${\bf q}$.

%\subsection{Arbitrariness in the gradient-mediated couplings}
\subsection{\black{Covariance principle}}
\label{sec_arbcoeff}

As expressed via Eqs.~(\ref{cfg}), it might appear
%at first sight
%seem reasonable to conclude 
that the three dispersion coefficients $C$, $f$ and $G$ 
%So far, most authors have implicitly assumed that all coefficients entering 
%Eq.~(\ref{free}) 
are well-defined (and measurable) physical properties of the 
crystal; most authors have indeed used such
an assumption in the past, either implicitly or explicitly. 
%
%In the following we shall demonstrate that, 
And yet, while $C$ is indeed a well-defined crystal property,
neither $f$ (as pointed out in earlier works~\cite{Hong-13})
or $G$ (as we shall demonstrate in the following) are; 
on the contrary, they both suffer from an unavoidable arbitrariness.
%an established and well-defined crystal property, neither $f$ or $G$ are, as
%they suffer from an unavoidable arbitrariness.
%
We anticipate that their ambiguity is \emph{physical}, i.e., it is 
not specific to the method one uses to calculate the coefficients 
within microscopic theory, and is directly related to the paradox 
of Fig.~\ref{cartoon}.
%; and
%it has profound consequences for the interpretation of the results.

%Given that the definition of the acoustic mode displacement is unique, and
%the long-wave expaansion of the transverse components of the force-constant matrix
%is well defined, the source of arbitrariness must be rooted in the definition 
%of the $|P_y\rangle$ vector.
%
After a quick glance at Eq.~(\ref{cfg}), it is not difficult to see where this 
arbitrariness may come from. Of the two ingredients that enter the definition
of $C$, $f$ and $G$, the $\Phi^{(2,xx)}$ matrix is well defined, as it directly 
emerges from a long-wave expansion of the force constants. 
As we have anticipated earlier, however, 
the basis vector $p^{(y)}_\kappa$
is only defined modulo a rigid displacement of the whole 
lattice (Fig.~\ref{cartoon}).
In other words, we can always replace $p^{(y)}_\kappa$ with any vector that 
differs from $p^{(y)}_\kappa$ by a $\kappa$-independent constant.
%, or in other words, by an arbitrary 
%constant times the .
%
There is no fundamental symmetry principle
that favors one choice over the other -- it is entirely a matter of convention.
Let's see what happens if we operate such a transformation, by defining a new 
\black{basis} vector as
\black{\begin{equation}
\label{ppr}
{p^{(y)}_\kappa}' = p^{(y)}_\kappa + \lambda.
\end{equation}}
Evidently, the homogeneous coupling coefficients are unaffected by such a
transformation, since the energetics of the uniform phase is insensitive
to the choice of the origin.
This is not the case for the gradient coefficients, whose transformation 
rules can be straightforwardly derived by plugging Eq.~(\ref{ppr}) into
Eq.~(\ref{cfg}),
\begin{subequations}
\label{fgpr}
\begin{align}
f' =& f + \lambda C, \\ 
G' =& G + 2 \lambda f + \lambda^2 C.
\end{align}
\end{subequations}
The fact that both the flexoelectric and gradient coefficients depend on 
an arbitrary constant, $\lambda$, 
%might appear surprising at first sight, but it's 
is actually easy to rationalize on elementary physical grounds.
Introducing a shift in the polar distortion is harmless in the homogeneous 
case, but the gradient of $P$ comes with an extra strain field,
which contributes to both the
%an additional 
elastic and flexoelectric terms in the free energy.

One might wonder, at this point, whether Eq.~(\ref{free}) can be trusted at all,
given the aforementioned arbitrariness. 
To answer this question, it is useful to understand the impact of 
Eq.~(\ref{ppr}) and Eq.~(\ref{fgpr}) on the domain wall solution.
%
%Remarkably, 
Crucially, the renormalized gradient coefficient $\widetilde{G}$ remains
unchanged, 
\begin{equation}
\black{\widetilde{G}' = \widetilde{G},}
\end{equation}
as the respective $\lambda$-dependent contributions to $f$ and $G$
exactly cancel out.
This means that all \emph{physically measurable} properties of the domain wall, 
i.e., the thickness $\xi$ and the energy $W$, are well defined regardless of
the specific convention that we choose for $p^{(y)}_\kappa$; the equilibrium solution
for $P(x)$ is also unaffected.
The only feature that changes with $\lambda$ is the equilibrium solution for the
elastic displacement field, and hence the \emph{local strain},
\begin{equation}
\black{u'(x)  =-\frac{f'}{C} P(x) 
  = u(x) - \lambda  P(x).}
\end{equation}
This result looks, \black{at first sight, surprising}: by modifying the definition of the
polar distortion we have obtained the same profile for $P(x)$, exactly the
same energy, but a different solution for the strain field.
The solution to this puzzle resides in Eq.~(\ref{micro}), which is
our gateway from the continuum solution back to the microscopics. 
And indeed, one can quickly verify that the change in the strain field 
exactly cancels out with the change in the displacements that are 
associated with the redefinition of $p^{(y)}_\kappa$, leaving the
equilibrium solution for the individual atomic displacements 
well-defined (that is, $\lambda$-independent) and unique.
This provides, in a nutshell, the solution to the paradox of Fig.~\ref{cartoon},
and constitutes one of our main formal results.

\black{That Eq.~(\ref{free}) behaves this way is not a coincidence, but
rather the consequence of a more general \emph{covariance principle}.
Suppose we operate the following transformation of the fields and the 
distortion vectors,
\begin{subequations}
\label{covariance}
\begin{align}
{p^{(\alpha)}_\kappa}' = & {p^{(\alpha)}_\kappa} + \lambda, \\
{\bf u}'({\bf r}) = & {\bf u}({\bf r}) - \lambda {\bf P}({\bf r}),
\end{align}
\end{subequations}
where $\lambda$ is an arbitrary dimensionless scalar.
The atomic distortions associated with the displacement and polarization fields 
%${\bf P}'({\bf r})$ and ${\bf u}'({\bf r})$ 
via Eq.~(\ref{basis2}) are manifestly invariant with respect to Eq.~(\ref{covariance}).
%such a 
%transformation.
%
This means that the original and primed quantities
refer to the same configuration of the system, i.e., they are
physically equivalent macroscopic representations of the same distorted
crystal structure.
%undistinguishable from the point of view of the microscopics.
%Symmetry principles alone cannot be used to favor $p^{(\alpha)}_\kappa$
%over ${p^{(\alpha)}_\kappa}'$, as both are $T_{1u}$ distortion modes;
%also, because of translational invariance, $p^{(\alpha)}_\kappa$
%and ${p^{(\alpha)}_\kappa}'$ describe equivalent directions in 
%the configuration space of the primitive cell.
%
It is natural then to require \emph{a priori} from any continuum functional
of ${\bf u}$ and ${\bf P}$ to be covariant with respect to the choice of $\lambda$,
i.e. that it transforms as %covariant (in a sense to be specified shortly)
Eq.~(\ref{covariance}).
The results of this Section demonstrate that Eq.~(\ref{free}) 
complies with such a requirement, % for this principle to hold,
provided that both flexoelectricity and polarization gradient coefficients 
are consistently calculated.}
%in any continuum description of inhomogeneous ferroelectric 
%structures.}

\subsection{Converse mapping to the continuum fields}
%: arbitrariness in the 
%            local strain}

\label{sec:converse}

It is ironic, in light of these results, to realize that macroscopic
theory is far better suited to predicting equilibrium atomic positions
rather than ``traditional'' macroscopic quantities, such as the strain.
To rationalize such an outcome, and get convinced
that ``it cannot be otherwise'', 
%a few more words of comments 
%on its implications are in order.
%
%In this context, 
it is illuminating to consider the \emph{converse}
mapping between continuum fields and microscopics, i.e., the procedure
that allows one to extract the values of ${\bf P}({\bf r})$ and 
${\bf u}({\bf r})$ given a distorted configuration of the crystal.
%
%We shall establish such a mapping in the following.
%
%This is what we shall indeed prove in the following paragraphs.
%
%(Establish the reverse mapping....)
%
We shall follow the same two-step procedure as in Section~\ref{sec:direct},
but taken in reverse order:
%
%Consider a crystal with more than one atom in the primitive 
%basis, which we assume to be distorted from its reference high-symmetry 
%structure by a displacement pattern that is not cell-periodic, 
%but mesoscopically modulated in space (that is, the distortion 
%amplitude varies on a length scale that is much larger than the
%interatomic spacings).
%
%Converting the atomic distortions into a mechanical 
%displacement field involves two separate steps: 
(i) transform the discrete sublattice distortions into continuum 
functions of all space; (ii) perform a local projection of the individual
atomic displacements onto the subspace spanned by \black{the active lattice 
modes, i.e., those associated to the fields 
%distortion patterns, 
${\bf v}_{\nu}({\bf r})$ via the transformation matrix ${\bf T}$}. %and  $|P_\alpha\rangle$.
%to obtain the vector fields of interest.

Step (i) does not involve any ambiguity as long as the 
atomic displacement pattern is mesoscopic in nature 
(i.e., the distortion amplitudes vary on a length scale 
that is much larger than the interatomic spacings).
Such an assumption implies that, if we express the distortion in 
reciprocal space, all phonon amplitudes, $u_{\kappa \alpha}({\bf q})$, vanish 
at the zone boundary.
This is a sufficient condition for the Fourier continuation of the
atomic displacements from discrete to continuum,
\begin{equation}
u^l_{\kappa \alpha} = u_{\kappa \alpha}({\bf R}^{(0)}_{l\kappa}) \rightarrow u_{\kappa \alpha}({\bf r})
\end{equation} 
to be uniquely defined.~\cite{Stengel-16}
%
%Of course, $u_{\kappa \alpha}({\bf q})$ is periodic, and therefore can 
%be regarded as a sum over the ${\bf G}$ vectors of $\bar{u}_{\kappa \alpha}({\bf q-G})$, where
%$\bar{u}_{\kappa \alpha}$ is equal to $u_{\kappa \alpha}$ within the first Brillouin zone, 
%and zero outside it.
%
%Since the repeated images of $\bar{u}_{\kappa \alpha}({\bf q})$ do not overlap, this leaves 
%$\bar{u}_{\kappa \alpha}({\bf q})$ uniquely defined; its backward Fourier transform readily
%yields the sought-after continuum field.
%
(Whenever the above condition breaks down one can still extract  
continuum fields from the atomistics by applying standard
%an appropriate Fourier 
%filter to the reciprocal-space phonon amplitudes. 
%Such a procedure is commonly known as
``macroscopic averaging''~\cite{baldereschi-88,junquera-07} techniques.) %within the first-principles community. 
%
%Strictly speaking, the result is no longer unambiguous;~\cite{junquera-07}
%but generally 
%depends on the choice of the Fourier filter~\cite{junquera-07}. 
%\st{This is, however, a regime where the continuum approximation is no longer 
%justified, and therefore falls outside the scopes of the present discussion.}
%)

%in the majority of
%practical cases, however, the dependence is rather mild, and doesn't hinder the
%correct interpretation of the results. %~\cite{deadlayer1,deadlayer2})

Regarding step (ii), we shall define the continuum fields by
inverting Eq.~(\ref{basis}),
%as follows,
\begin{equation}
\label{vk}
\black{v_{\nu\beta}({\bf r}) =  \sum_{\kappa \alpha} %\langle \tilde{v}_\alpha | \kappa \beta \rangle 
%\tilde{v}^{(\beta)}_{\kappa \alpha}
u_{\kappa \alpha} ({\bf r}) \tilde{T}_{\kappa \alpha,\nu\beta},}
\end{equation}
\black{where %$v$ stands as usual for $u$ or $P$, and 
$\tilde{T}_{\kappa \alpha,\nu\beta}$
is the converse transformation matrix, which satisfies the 
condition $\tilde{\bf T} {\bf T} = {\bf I}$.
Similarly to the direct one, $\tilde{\bf T}$ is diagonal 
on the Cartesian indices, 
$\tilde{T}_{\kappa \alpha,\nu\beta} = \delta_{\alpha\beta} \tilde{T}_{\kappa \alpha,\nu\alpha}$.
The columns referring to $\nu=u,P$, which we indicate as
$\tilde{u}^{(\alpha)}_{\kappa} = \tilde{T}_{\kappa \alpha,u\alpha}$
and $\tilde{p}^{(\alpha)}_{\kappa} = \tilde{T}_{\kappa \alpha,P\alpha}$
%$\langle \tilde{v}_\alpha |$ 
are the \emph{duals} to the direct
basis vectors, respectively ${u}^{(\beta)}_{\kappa}$ and ${p}^{(\beta)}_{\kappa}$},
% $|u_\alpha \rangle$ and $|P_\alpha \rangle$ 
that we have introduced in Section~\ref{sec:direct}.
The basic requirement on the direct and dual vectors is that they form 
an orthonormal set, 
\black{\begin{equation}
\label{orthon}
\begin{split}
      \sum_\kappa \tilde{u}^{(\alpha)}_{\kappa}{u}^{(\beta)}_{\kappa} =& 
      \sum_\kappa \tilde{p}^{(\alpha)}_{\kappa}{p}^{(\beta)}_{\kappa} = \delta_{\alpha \beta}, \\ 
      \sum_\kappa \tilde{u}^{(\alpha)}_{\kappa}{p}^{(\beta)}_{\kappa} =& 
      \sum_\kappa \tilde{p}^{(\alpha)}_{\kappa}{u}^{(\beta)}_{\kappa} = 0.
%\langle \tilde{u}_\alpha |u_\beta \rangle = & \langle \tilde{P}_\alpha |P_\beta \rangle = \delta_{\alpha \beta}, \\ 
%\langle \tilde{P}_\alpha |u_\beta \rangle = & \langle \tilde{u}_\alpha |P_\beta \rangle = 0.
\end{split}
\end{equation}}
This is a necessary condition to ensure consistency, e.g., that a subsequent 
application of the direct and converse mapping recovers the initial values of the 
continuum fields.
The most general choice that satisfies these constraints consists in 
introducing a set of sublattice-dependent weights, $w_\kappa$, whose sum 
is unity, $\sum_\kappa w_\kappa = 1$.
Then, we define the duals as
\begin{subequations}
\begin{align}
\tilde{u}^{(\beta)}_{\kappa} =& w_\kappa, %\langle \tilde{u}_\alpha | \kappa \beta \rangle =& w_\kappa \delta_{\alpha \beta}, 
 \label{dualu} \\ 
% \langle \tilde{P}_\alpha | \kappa \beta \rangle =& w_\kappa p_\kappa \delta_{\alpha \beta}. 
\tilde{p}^{(\beta)}_{\kappa} =& w_\kappa {p}^{(\beta)}_{\kappa}.  \label{dualp}
\end{align}
\end{subequations}
This way, the orthonormality of the elastic displacement vectors is 
enforced by construction, while the remainder of Eq.~(\ref{orthon})
leads to the following two conditions on $p_\kappa^{(\beta)}$,
\begin{subequations}
\begin{align}
\sum_\kappa w_\kappa p^{(\beta)}_\kappa =& 0, \label{conda} \\
\sum_\kappa w_\kappa [p^{(\beta)}_\kappa]^2 =& 1. \label{condb}
\end{align}
\end{subequations}

Eq.~(\ref{conda}) is a ``hard'' requirement on $p_\kappa$, and
must always be enforced after some choice of weights is made.
%
%We are only left with explaining how the choice of the weights 
%affects the definition of the polar mode $|P_\alpha\rangle$.
%, though;
%we shall do it now.
%
This condition lifts the indeterminacy of the polar distortion
vector that we illustrated in Fig.~\ref{cartoon}, and 
clarifies the role of $w_\kappa$ in subtracting the
(weighted) average displacement of the cell from the polar mode.
Doing so is consistent with physical intuition: 
the polarization, by its nature, is a distortion
of the lattice that does not move the unit cell of 
the crystal as a whole.
%
%The exact meaning of the condition ``does not move the unit 
%cell'' is formalized by Eq.~(\ref{conda}), which emerges
%here as a necessary condition for the two-way mapping between 
%macro and micro to be mathematically sound.
%\begin{equation}
%\label{wp}
%\sum_\kappa w_\kappa \langle \kappa \alpha| P_\beta \rangle = 0.
%\end{equation}
%
\black{Thus, the translational freedom that we have described 
in the earlier Sections can be equivalently expressed as 
a \emph{weight freedom} in the converse mapping to the macroscopics,
which provides an even more direct connection to the theory of 
Ref.~\onlinecite{Hong-13}.}

Eq.~(\ref{condb}), on the other hand, is a consequence of Eq.~(\ref{dualp}),
which is to some extent arbitrary. Indeed, one can always multiply Eq.~(\ref{dualp})
by a constant factor;
%$\langle P_\alpha | \kappa \beta \rangle = \tilde{p}_\kappa \delta_{\alpha \beta}$,
%with the only condition that the components of the dual vector sum up to zero, 
%$\sum_\kappa \tilde{p}_\kappa = 0$. 
such freedom boils down to the choice of units that we use to measure the polar 
distortion amplitude. 
(For example, one could require $P_0$ to coincide with the spontaneous polarization 
of the ferroelectric crystal, as customary in macroscopic theories.)
%one could set $\tilde{p}_\kappa$ to the Born effective charges of the 
%individual atoms: their sum must vanish due to the acoustic sum rule, and 
%$P(x)$ would then be measured in units of polarization, 
%
%(.)
%
The present convention, which consists in measuring $P(x)$ in length units, 
has the drawback that the normalization condition [Eq.~\ref{condb}], and
hence the values of all coefficients of Eq.~(\ref{free}), depends
on the choice of weights. Still, we shall prefer it here because it bears
a direct formal link to the eigenvectors of the dynamical matrix (see Appendix~\ref{sec_eigrep}),
and for consistency with earlier works.~\cite{Stengel-16,Schiaffino-17,Zabalo-21}.

Eq.~(\ref{conda}) and~(\ref{condb}), together with the prescriptions of Section~\ref{sec:direct},
yield a well-defined procedure to construct the eigendisplacement vectors, and hence
the model parameters, given a set of weights $w_\kappa$. 
Starting from the atomic distortions in the relaxed ferroelectric structure with the 
polarization oriented along $\beta$, $d^{(\beta)}_\kappa$, we 
first of all enforce Eq.~(\ref{conda}) via
\begin{equation}
\bar{d}^{(\beta)}_\kappa = d^{(\beta)}_\kappa - \sum_\kappa w_\kappa d^{(\beta)}_\kappa.
\end{equation}
Then, we define the amplitude of the 
spontaneous distortion as 
\begin{equation}
P_0 = \sqrt{\sum_\kappa w_\kappa [\bar{d}^{(\beta)}_\kappa]^2}.
\end{equation}
Finally, we enforce Eq.~(\ref{condb}) by defining the 
dimensionless eigendisplacement vector as $p^{(\beta)}_\kappa = \bar{d}^{(\beta)}_\kappa / P_0$.
%To construct the polar eigendisplacement vector $p_\kappa$, 
%we first of all remove the center of mass from $d_\kappa$ via Eq.~(\ref{xxx})
%and Eq.~(\ref{yyy}). 
%
%We normalize the resulting $\tilde{d}_\kappa= d_\kappa - \lambda$ 
%by 
%%
%Then, Eq.~(\ref{zzz}) uniquely defines the dimensionless 
%The result manifestly depends on the (arbitrary) choice of the 
%weights in a way that reflects the arbitrariness 
%; we shall discuss the physical implications of this outcome
%in the following subsection.

%\black{STOPPED HERE.}

\begin{table*}%[!b]
\begin{center}
\begin{tabular}{cddddddd}
\hline \hline
& \multicolumn{1}{c}{$a_0$}
& \multicolumn{1}{c}{$d_{A}$}
& \multicolumn{1}{c}{$d_{B}$}
& \multicolumn{1}{c}{$d_{{\rm O}_{1}}$}
& \multicolumn{1}{c}{$d_{{\rm O}_{2,3}}$}
& \multicolumn{1}{c}{$\Delta E$}
& \multicolumn{1}{c}{$P$} \\
\hline
BaTiO$_3$ & 3.935 & +0.0186 & +0.\black{0}538 & -0.0359 & -0.0183 &   -2.87 & 0.188 \\
BaTiO$_3^{(*)}$
          & 3.904 & +0.0087 & +0.0237 & -0.0117 & -0.0064 &  -0.083 & 0.078  \\
CaTiO$_3$ & 3.799 & +0.2369 & +0.0584 & -0.0432 & -0.1261 &  -36.72 & 0.508 \\
 KNbO$_3$ & 3.947 & +0.0244 & +0.0553 & -0.0229 & -0.0287 &   -3.77 & 0.205 \\
NaNbO$_3$ & 3.907 & +0.2097 & +0.0498 & -0.0480 & -0.1057 &  -18.75 & 0.359 \\
PbTiO$_3$ & 3.879 & +0.1894 & +0.0793 & -0.0524 & -0.1081 &  -30.56 & 0.578 \\
PbZrO$_3$ & 4.102 & +0.3674 & +0.0716 & -0.0241 & -0.2075 & -187.50 & 0.647 \\
\hline \hline
\end{tabular}
\caption{
Data computed using DFT for six perovskite oxides:
%(the second entry refers to BaTiO$_3$ under pressure):
simple-cubic lattice parameter, $a_0$ (in~\AA);
displacements (\black{in~\AA}) of the atoms from
their high-symmetry positions in the distorted tetragonal structure
at fixed lattice constants $a_0$, $d_{\kappa}$,
for the $A$ cation ($d_{A}$), $B$ cation ($d_B$),
apical anion ($d_{{\rm O}_1}$), and
equatorial anions ($d_{{\rm O}_{2,3}}$);
energy of this relaxed tetragonal
configuration with respect to the simple cubic one (in meV/f.u.);
and polarization of the tetragonal phase (in C/m$^2$).
(*): calculations of BaTiO$_3$ under hydrostatic pressure, see
Sec.~\ref{sec:pressure}.
}
\label{tab_fiveatomcells}
\end{center}
\end{table*}

\subsection{Arbitrariness of the strain field}

With the above derivations, we have established
the continuum displacement field
as a weighted average over all sublattices,
\begin{equation}
\label{wk}
{\bf u}({\bf r}) = \sum_\kappa w_\kappa {\bf u}_\kappa ({\bf r}).
\end{equation}
%where $w_\kappa \geq 0$ are the weights. 
%
The issue with this formula, which is otherwise rather trivial, is the fact that
the weights are completely arbitrary. There may be, of course, some choices that 
are preferrable over others, for different reasons.
%but the key point is that there is no \emph{fundamentally} right (or wrong) 
%choice.
%
%
Several authors~\cite{Wang-20,Stengel-16}, for example, advocate the use of 
the physical masses of the atoms as weights; this is convenient for dynamical 
problems, where masses indeed play a role, and provides the physically intuitive
interpretation of the displacement field as the displacement of the local 
center of mass.
%
%Yet, other choices are equally valid: m
Then, microscopists routinely use
the positions of the heaviest ions to define the local strain, as
they correspond to the brightest spots in the images; this implies
setting their weight to unity, and the others to zero.
Simply taking the average displacement of the cell (with equal weights) is 
not uncommon, either.

The key point is that there is no \emph{fundamentally} right (or wrong) 
choice: since spatial inversion is broken in the polar structure, 
within the bulk domains the relation between the cell origin and the 
atomic positions cannot be fixed by symmetry (see Fig.~\ref{cartoon}).
Yet, the ``covariance'' of Eq.~(\ref{free}) with respect to the weight 
arbitrariness guarantees that the physics is uniquely described, even if
the strain field (and hence the net elastic offset across the wall,
$\Delta u$) depends on such choice.
Note that this result, which has been established here for a static 
domain-wall structure, holds in full generality: In Appendix~\ref{sec_dyn} 
we generalize it to the time-dependent regime, and use it to reconcile 
the existing controversies around the so-called ``dynamical flexoelectric effect''.

%One may wonder whether this invariance principle breaks down in
%dynamical problems, where masses do play a role. 
% regardless of 
%this inherent arbitrariness.
%
%One must keep in mind

% all of them agree in cases well the strain is locally uniform,
%and generally disagree otherwise.

%We are only left with showing that 
An important consequence of the formalism developed here is that
%
%The derivations of the previous section demonstrate that such 
%arbitrary choice of the atomic weights coincides with the 
%arbitrariness in the definition of the polarization vector
%(and hence of the model coefficients) that we have discussed 
%earlier. 
%
%Thus, it turns out that 
the definition of strain and 
polarization are intimately related: they are
both bound, respectively via Eq.~(\ref{wk}) 
and Eq.~(\ref{conda}), to the same weight choice ambiguity.
And indeed, \black{Eq.~(\ref{covariance}) shows} that ${\bf u}({\bf r})$ 
\emph{is ambiguous only \black{in presence of a spatially nonuniform
polarization,}
%the strain or 
%the polarization are nonuniform in space,
%and a uniform polarization is simultaneously present, 
and is uniquely defined otherwise.}
%
% is not immediately obvious: Eq.~(\ref{ppr}) describes an 
%ambiguity in the definition of the polar distortion, while
%Eq.~(\ref{wk}) points to an ambiguity in the definition of
%the mechanical displacement field.
%
%The two are intimately related, though: it turns out that
%}
%
\black{To see this, recall that the strain is defined as the first 
gradient of the displacement field. If ${\bf P}({\bf r})$
vanishes or is constant over space, Eq.~(\ref{covariance})
yields $\partial u_\alpha / \partial r_\beta = \partial u'_\alpha / \partial r_\beta$,
independent of $\lambda$, i.e., the strain becomes a well-defined quantity.
%the transformation law
%for ${\bf u}'({\bf r})$ 
%
%Eq.~(\ref{ub}) and Eq.~(\ref{micro}): 
%in absence of ${\bf P}({\bf r})$, the atomic displacements 
%corresponding to a uniform strain field ${\bf u}({\bf r})$ are 
%always given by an equal displacement of all sublattices,
%\begin{equation}
%${\bf u}_\kappa ({\bf r}) = {\bf u}({\bf r}).$
%\end{equation}
%
This is manifestly consistent with Eq.~(\ref{wk}): if 
all the ${\bf u}_\kappa ({\bf r})$ are equal modulo a constant
(which is true if the polarization is uniform),
their spatial gradients coincide; then, any choice of
the weights yields the same result for the strain.}
%
%Also, in absence of ${\bf P}$, the free energy yields a
%unique solution for the strain field since the elastic tensor
%is a well-defined material property.
%
%Of course, whenever \emph{both} strain and polarization are uniform in
%space, the strain is a well-defined quantity.

We have achieved, therefore, a complete physical picture.
There are three, at first sight unrelated, ambiguities in 
the mapping from continuum to atomistics and vice versa,
and concern: (i) the definition of the local strain; (ii) 
the definition of the polar distortion; (iii) the definition
of the flexocoupling and gradient coefficients in the free energy.
We have shown that (i--iii) share the same formal root, and
can be expressed as a freedom in the choice of a set of 
atomic weights, $w_\kappa$.
This choice should be made once and for all at the beginning, 
and consistently respected throughout the calculation of all 
free-energy coefficients; then, the physical answers that we 
extract from Eq.~(\ref{free}) should not depend on the specific 
set of $w_\kappa$ that we use. 
The variational solution of the continuum differential equations
does depend, in general, on $w_\kappa$, but it must be this way:
if we are asking, for example, ``what is the local displacement 
of the cell at the point ${\bf r}$'', the answer inevitably 
depends on how we define such a displacement via Eq.~(\ref{wk}).
Similar considerations hold whenever we use the information
on ${\bf u}({\bf r})$, extracted from experimental or theoretical
domain-wall structures via Eq.~(\ref{wk}), to 
estimate the flexocoupling coefficient by inverting Eq.~(\ref{offset}).

\begin{figure*}%[!b]
\centering
\includegraphics[width=40mm,angle=-90]{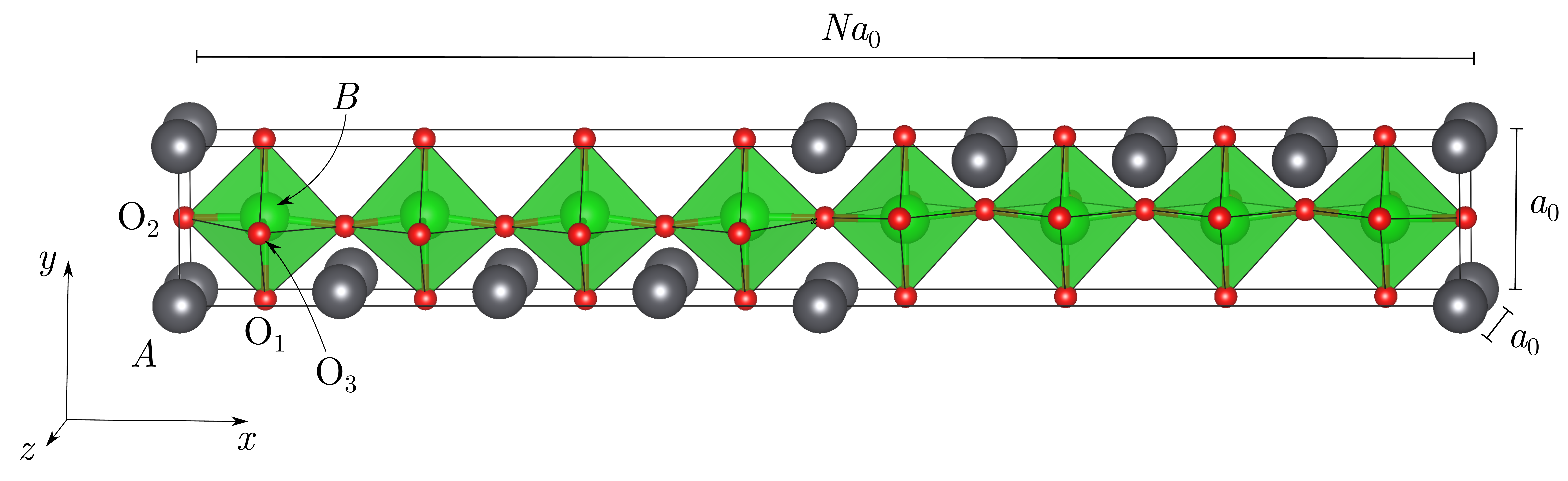}
\caption{(Color online.)
Example of unit cell used in some of our full-DFT calculations
(here, $N=8$, and
domains with opposite orientations give rise to an $A$O-type domain wall).
}
\label{fig_supercell}
\end{figure*}

%%%%%%%%%%%%%%%%%%%%%%%%%%%%%%%%%%%%%%%%%%%%%%%%%%%%%%%%%%%%%%%%%%%%%%%%%%%%%%%%

\section{Results}

\label{sec:results}

\subsection{Computational parameters}
\label{sec_comp}

%In order to compute the coefficients of the free energy expansion of Equation
%(\ref{renorm}) 
Our calculations are performed in the framework of DFT 
as implemented in the ``in-house'' {\sc Lautrec} code \cite{lautrec}.
We use the local-density approximation \cite{perdew/wang:1992},
the projector augmented wave method (PAW) \cite{bloechl:1994},
and a plane-wave basis set with a kinetic energy cutoff of 50 Ry
in all our calculations.
The PAW datasets are generated by treating the 
following orbitals as valence electrons:
Ba($5s^2 5p^6 6s^2$), 
Ca($3s^2 3p^6 4s^2$),
K($3s^2 3p^6 4s^1$),
Na($2s^2 2p^6 3s^1$),
Nb($4s^2 4p^6 5s^2 4d^3$),
O($2s^2 2p^4$),
Pb($6s^2 5d^{10} 6p^2$),
Sr($4s^2 4p^6 5s^2$),
Ti($3s^2 3p^6 4s^2 3d^2$),
and 
Zr($4s^2 4p^6 5s^2 4d^2$).
In all calculations, the Brillouin zone is sampled using 
Monkhorst-Pack \cite{monkhorst-76} meshes that are equivalent (or
better) to an $8 \times 8 \times 8$ $k$-point grid in the 
five-atom bulk cell.

For each of six $AB$O$_3$ perovskites (BaTiO$_3$, CaTiO$_3$, KNbO$_3$, 
NaNbO$_3$, PbTiO$_3$, and PbZrO$_3$) we first calculate the equilibrium 
lattice parameter $a_0$ of the cubic reference structure by fitting
the energy to the Murnaghan equation of state. (The resulting values are listed in Table
\ref{tab_fiveatomcells}). 
Using the corresponding $a_0 \times a_0 \times a_0$ five-atom cells, we then
displace the atoms along one of the main axes
and re-optimize their positions, leading to
tetragonal configurations with lower energy. 
The displacements from the cubic positions, $d_\kappa$, 
and the energy differences, $\Delta E$, and spontaneous 
polarization $P$ (calculated via the method of
Ref.~\cite{Stengel-05}) are reported in Table \ref{tab_fiveatomcells}.
%
%The simplest option consists in requiring that $F_0$ coincide with the
%calculated energy difference between the reference cubic phase and the 
%ferroelectric state, $\Delta E$, which leads to
\black{Based on the calculated values of $\Delta E$ and 
equilibrium polar distortion amplitude $P_0$ we then
calculate $A$ and $B$ as
\begin{eqnarray}
A  =   \frac{4\Delta E}{\Omega P_0^2}, \qquad 
B =  -\frac{4\Delta E}{\Omega P_0^4}.
\end{eqnarray}
}

In order to calculate the second-order term in the long-wave 
expansion of the force-constant matrix,  Eq.~(\ref{lw}), 
we use the real-space supercell approach of Ref.~\onlinecite{Hong-13}.
This implies calculating the second-order moments of the
interatomic force constants according to
\begin{equation}
\label{sums}
\Phi^{(2,xx)}_{\kappa y, \kappa' y} = \sum_l \Phi^{l}_{\kappa y, \kappa' y} ({\bf R}_l + \bm{\tau}_{\kappa'} - \bm{\tau}_\kappa)_x^2.
\end{equation}
[Due to inversion symmetry, displacements of the atoms along $y$ do not generate 
electric fields along $x$; this implies that the interatomic forces decay exponentially 
with distance, and the lattice sums in Eq.~(\ref{sums}) converge to a unique, well defined
value.]
To compute $\Phi^{l}_{\kappa y, \kappa' y}$,
%of Equation \ref{sums}, 
we carry out calculations in which we displace one atom
at the time by 0.005 a.u. along $y$ and extract the resulting forces.
In practice, we use a $8 a_0 \times a_0 \times a_0$ supercell, with
the same geometry as in Figure \ref{fig_supercell},
except that we use the centrosymmetric paraelectric structure as reference.
(The resulting matrix elements are given in Appendix A.)

%but only one
%atom at a time was displaced from the high-symmetry positions of the 
%simple-cubic structure) 
%and $1 \times 8 \times 8$ Monkhorst-Pack meshes.

%\black{For simplicity, we shall focus our discussion to systems where macroscopic
%electric fields play no role; this is the case, for example, of a 180 degrees
%domain wall in a perovskite crystal with a cubic high-symmetry reference phase.
%
%We shall assume a (100)-oriented wall, so the zone-center polar mode $|P_\alpha \rangle$ 
%only contains atomic displacements along the direction $\alpha$.
%Then, we only need to consider elements of the type $\Phi^{(2,xx)}_{\kappa y, \kappa' y}$,
%which can be equivalently defined as real-space moments of the interatomic force constants,
%
%}

Finally, to validate the model results against full DFT calculations we
prepare two domains with opposite polarization in a long (100)-oriented 
supercell (as in Figure \ref{fig_supercell}, for different values of $N$), and
we allow the atoms to relax along $y$ till the forces on them are
negligible.
We consider both $A$O-centered and $B$O$_2$-centered wall types.
Note that we neglect octahedra rotations, strain relaxations, or other wall 
orientations that might result in energetically more favorable structures.
Our main goal here is testing the continuum approximation on a 
minimal Landau model of a ferroelectric wall, and discussing the 
subtleties related to the \black{treatment} of \black{gradient} 
effects. 
In this sense, incorporating additional degrees of freedom 
to achieve a more realistic picture would have constituted an
unnecessary complication.
In some members of our materials set such a simplified model does
not yield a physically meaningful description of the bulk or domain 
wall structure (or both). 
For this reason, we shall primarily focus
our attention on BaTiO$_3$, PbTiO$_3$ and KNbO$_3$,
and present the data on other materials for
comparison purposes and future reference.

\subsection{DFT calculations of domain walls}

\begin{figure*}
\centering
\includegraphics[width=5in,angle=0]{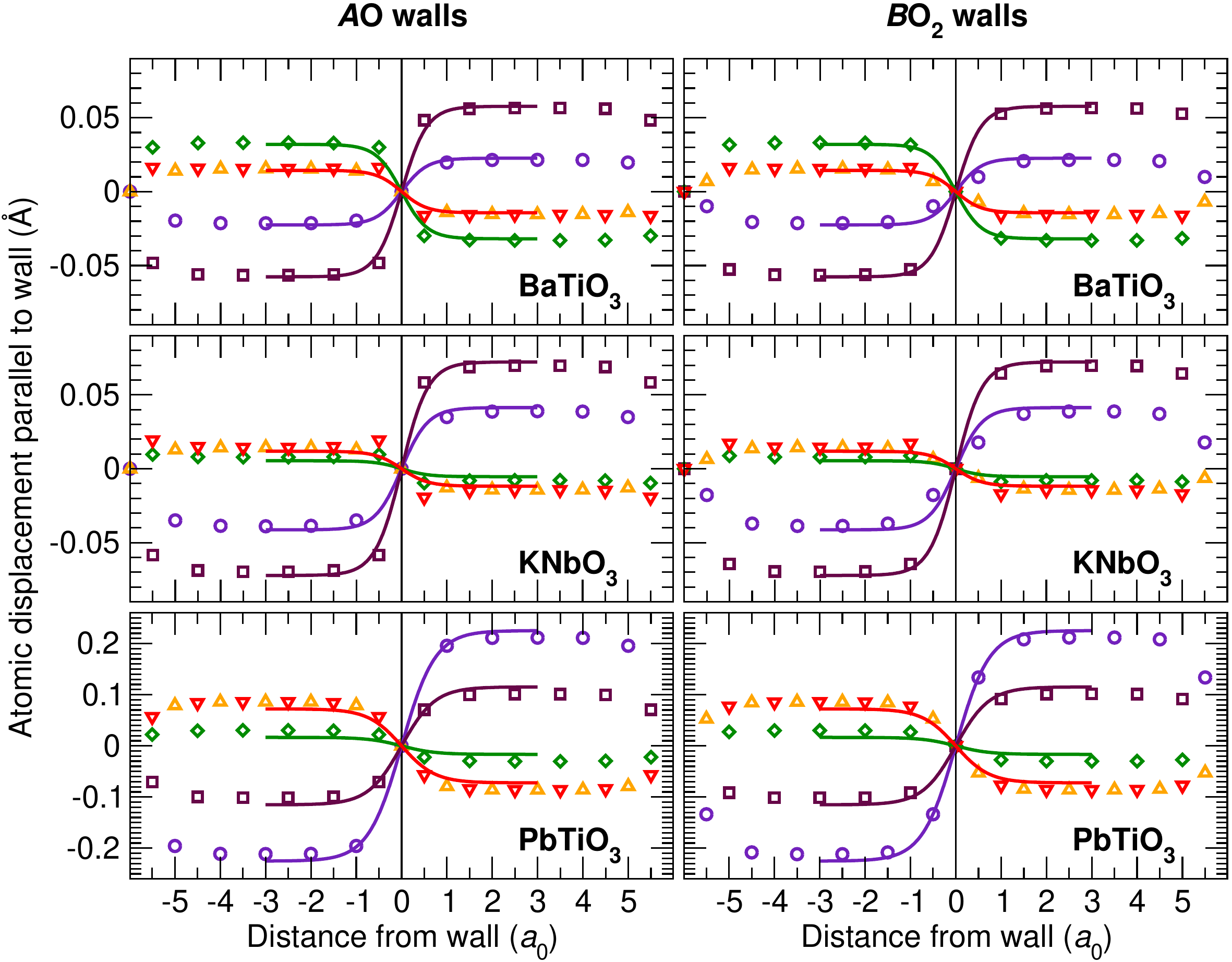}
\caption{(Color online.) 
\black{Empty symbols: Relaxed displacements of the atoms in the 
direction parallel to the domain wall as a function
of the normal coordinate, extracted from full DFT calculations
(circles, squares, diamonds, and triangles represent, respectively,
$A$, $B$, apical O, and equatorial O atoms).
Continuous curves: atomic displacement
profiles predicted by the Landau model.}
\black{Two possible types of domain wall (either AO- or BO$_2$-centered) 
are shown for three of the six perovskites considered in this work.}
}
\label{fig_dwpattern}
\end{figure*}

We start by discussing our direct DFT calculation of the 
domain-wall structures.
Figure \ref{fig_dwpattern} shows the resulting atomic displacements from
the high symmetry positions.
Two features are common to all materials:
the domain walls are atomically thin,
and the atomic positions at the center of each domain depend only on the
material, and not on the type of wall ($A$O or $B$O$_2$).
%
%Figure \ref{fig_dwpattern} also shows that 
The six oxides considered here can be roughly classified into two categories:
those for which the relative
displacement between $A$ cations in adjacent domains is significantly larger
than the one between $B$ cations (CaTiO$_3$, NaNbO$_3$, PbTiO$_3$, and
PbZrO$_3$), and those for which this is not the case (BaTiO$_3$ and KNbO$_3$).
Such an outcome reflects the bulk distortion patterns quoted in Table 
\ref{tab_fiveatomcells}, which indeed shows that BaTiO$_3$ and KNbO$_3$
have similar properties, e.g., regarding the small displacements of their $A$
cations.
A similar classification also applies to the domain wall energies, listed 
in Table \ref{tab_prop}: for BaTiO$_3$ and KNbO$_3$
the domain wall energies are significantly smaller than in
other oxides (about one order of magnitude smaller than in 
PbTiO$_3$), with the $B$O$_2$-type wall energy approximately 50\% 
higher than the $A$O-type value. In other oxides the energies are 
larger, and for both types of wall they are within
25\% of each other. (The $B$O$_2$ type becomes favored for 
CaTiO$_3$ and PbZrO$_3$.) 
This picture is consistent with earlier calculations~\cite{Meyer2002PRB} in PbTiO$_3$
 and BaTiO$_3$, even though in our
calculations the relaxation of the cell parameters is not allowed.
% were carried out allowing
%for cell parameter optimization).
% 
%In line with the previous literature, we find that
%the domain wall energy in PbTiO$_3$ is around one order of magnitude larger
%than in BaTiO$_3$.

\begin{table}%[!b]
\begin{center}
\begin{tabular}{cdddc}
\hline \hline
& \multicolumn{3}{c}{$W$ (mJ/m$^2$)} 
& \multicolumn{1}{c}{$\quad \xi (\rm{\AA}) \quad$} \\
%& \multicolumn{3}{c}{$p_A$ (\AA)} 
& \multicolumn{1}{c}{DFT ($A$O)}
& \multicolumn{1}{c}{DFT ($B$O$_2$)} 
& \multicolumn{1}{c}{Eq.\ (\ref{eqdwe})} 
& \multicolumn{1}{c}{Eq.\ (\ref{eqxi})}
\\
%& \multicolumn{1}{c}{DFT ($A$O)} 
%& \multicolumn{1}{c}{DFT ($B$O$_2$)} 
%& \multicolumn{1}{c}{Landau Model} \\
\hline
BaTiO$_3$     &   2.95  &   4.59 &   4.43   &  2.204 \\ 
BaTiO$_3^{(*)}$ &   0.321 & 0.322  &   0.324  &  5.441 \\
CaTiO$_3$     &  63.2   &  60.8  &  70.5    &  2.463 \\ 
 KNbO$_3$     &   4.46  &   6.61 &   6.03   &  2.303 \\ 
NaNbO$_3$     &  26.6   &  32.7  &  34.3    &  2.557 \\ 
PbTiO$_3$     &  54.9   &  58.1  &  63.8    &  2.855 \\ 
PbZrO$_3$     & 200     & 133    & 252      &  2.170 \\ 
\hline \hline
\end{tabular}
\caption{
Domain wall properties for six perovskite oxides:
domain wall energy $W$ as computed using DFT (for the two possible
$A$O and $B$O$_2$ domain walls) and as computed using the Landau model;
and domain wall thickness $\xi$ as computed using the
Landau model.
% and displacement of the $A$ cation at
% the center of a domain, as computed using DFT (for the two possible
% $A$O and $B$O$_2$ domain walls) and as computed using the Landau model.
}
\label{tab_prop}
\end{center}
\end{table}

The atomic configurations and energies of some perovskite oxide domain walls
have been studied in the past using DFT-based methods.
Padilla, Zhong, and Vanderbilt \cite{Padilla1996PRB} carried out a pioneering
study on 180$^\circ$ domain walls in BaTiO$_3$ using an effective Hamiltonian
built from DFT results; they reported that the walls are atomically thin and
centered at the Ba atoms, consistent with our results,
and that the domain wall energies are of the order of 10 mJ/m$^2$.
Full DFT studies of 180$^\circ$ domain walls in PbTiO$_3$, 
first by P\"oykk\"o and Chadi\cite{Poykko1999APL}
and later by Meyer and Vanderbilt\cite{Meyer2002PRB},
reached the same conclusion regarding thickness, and predicted domain wall
energies of 100 to 200 mJ/m$^2$. In this case, the most favorable domain walls 
are found to be centered on the Pb atoms, which is again consistent with our
findings.
To the best of our knowledge, Ref.~\cite{Meyer2002PRB} was the first
to point out the geometrical offset of the atomic rows
between the oppositely polarized domains; we shall discuss this point
extensively in the next Section.
%These findings were later corroborated by  using similar
%methodology; they also investigated 90$^\circ$ domain walls, finding out that
%they have a much lower domain-wall energy of 35 mJ/m$2$.

\black{More recent studies have revealed that surprises may be in store 
even in systems that were hitherto believed to be simple and well understood.
A particularly illuminating example concerns the prediction of secondary Bloch-like
components in PbTiO$_3$.~\cite{Wojdel-14}
The contribution of secondary (or co-primary) antiferrodistortive modes (involving rotations
of the O$_6$ octahedra) to the domain wall energy and structure has also been studied in
selected cases.\cite{Lubk2009PRB,Dieguez2013PRB}
These works clearly indicate that our Eq.~(\ref{free}) is too simplified to 
provide a realistic picture in many materials; a follow-up work is currently 
under way to generalize our model to more complex geometries and boundary conditions.}
%Regarding perovskite oxides where  rotations are important,
%DFT-based studies have shown that the walls
%are still one or two unit cells in width, and that the matching of the
%octahedra at the wall strongly affects the value of the wall energy.
%
Although experimental probing of the structure and energetics of ferroelectric
domain walls is still challenging, \black{by now the characterization methods are 
mature enough to allow for a meaningful comparison to theoretical results.
For example, it is now widely accepted that} domain walls in perovskite 
oxides can be atomically thin (see, for example, Ref.\ 
\onlinecite{Evans2020PSR} and other references therein) as we have found here.

\begin{table}%[!b]
\begin{center}
\begin{tabular}{cddddd}
\hline \hline
& \multicolumn{1}{c}{$A$} 
& \multicolumn{1}{c}{$B$} 
& \multicolumn{1}{c}{$C$} 
& \multicolumn{1}{c}{$f$} 
& \multicolumn{1}{c}{$G$} \\
% & \multicolumn{1}{c}{$\tilde{G}$} \\
\hline
BaTiO$_3$ & -0.276  & 74.6  & 4.61  & -0.560 & 2.47  \\
BaTiO$_3^{(*)}$ & -0.0503  & 83.3   & 4.77   & -0.590 & 2.66   \\
CaTiO$_3$ & -0.219  &  3.29 & 3.47  & -0.861 & 2.59 \\ % & 0.878 \\
 KNbO$_3$ & -0.322  & 77.8  & 3.27  & -1.62  & 3.86  \\ % & 1.27  \\
NaNbO$_3$ & -0.135  &  2.65 & 2.56  & -1.36  & 2.30  \\ % & 0.634 \\
PbTiO$_3$ & -0.234  &  4.79 & 3.45  & -1.06  & 3.73  \\ % & 1.34  \\
PbZrO$_3$ & -0.365  &  2.26 & 2.09  & -0.386 & 3.14  \\ % & 1.43  \\
\hline
BaTiO$_3$ & -0.448  & 195   & 4.61  & -3.82  & 7.04  \\ 
BaTiO$_3^{(*)}$ & -0.0801 & 212     & 4.77   & -3.86   & 7.35    \\
CaTiO$_3$ & -0.219  &  3.30 & 3.47  & -2.28  & 3.87  \\ % & 0.88  \\
 KNbO$_3$ & -0.316  & 74.9  & 3.27  & -4.15  & 8.26  \\ % & 1.24  \\
NaNbO$_3$ & -0.215  &  6.72 & 2.56  & -2.60  & 5.14  \\ % & 1.01  \\
PbTiO$_3$ & -0.302  &  8.02 & 3.45  & -5.49  & 13.1  \\ % & 1.73  \\
PbZrO$_3$ & -0.431  &  3.14 & 2.09  & -2.75  & 7.24     % & 1.69 
\\
\hline \hline
\end{tabular}
\caption{
Coefficients of the Landau model of
Eq.~(\ref{free}) computed from DFT calculations. The
values are calculated by expressing both $u$ and $P$ fields in
Bohr units of length. Then, $A$ is in $10^{-3}$ Ha bohr$^{-5}$;
$B$ in $10^{-3}$ Ha bohr$^{-7}$; $f$, $G$ and $C$ are all in 
atomic units of pressure, i.e., $10^{-3}$ Ha bohr$^{-3}$.
Top block: unit weights are used. Bottom block: physical
atomic masses are used as weights.
}
\label{tab_coeffs}
\end{center}
\end{table}

%%% I THINK THIS WOULD BE BETTER SOMEWHERE ELSE (INTRO, CONCLUSIONS...)
%A natural question that arises on inspection of Figure \ref{fig_dwpattern}
%regards the amount of displacement of the atoms of the middle of a domain
%with respect to the ones in an adjacent opposite domain. 
%For example, do a particular set of ions remain at the same level on both
%sides? Or maybe some center of mass for a particularly simple choice of
%weights?
%The results of next section help to answer this kind of question.

\subsection{Landau model calculations of domain walls}

In this Section we develop Landau models for each of our six perovskite
oxides according to the guidelines specified earlier.
We start from the 
calculated distortion pattern ($d_\kappa$) and 
energy gain ($\Delta E$)  
of the tetragonal ferroelectric phase, 
as reported in Table \ref{tab_fiveatomcells}.
After processing the latter values via the procedure 
described in Sec.~\ref{sec:direct} and Sec.~\ref{sec:converse}
we readily obtain the values of $A$, $B$ and $p_\kappa$ for each 
bulk material.
The gradient-mediated coefficients of the model ($C$, $f$, $G$)
% and a combination
%of those, $\tilde{G}$) 
are then computed from $p_\kappa$ and the calculated 
$\Phi^{(2,xx)}_{\kappa y, \kappa' y}$ 
%from the elements of the 
%second-moment interatomic force-constant matrix 
via Eq.~(\ref{cfg}).
To illustrate our arguments, we shall use two 
different choices of weights for defining $p_\kappa$
(and hence the $A$, $B$, $f$ and $G$ parameters of the model),
by setting $w_\kappa$ either to equal values or to 
the physical masses of the atoms.
We shall compare the results and demonstrate their
mutual consistency in the following.

In Table \ref{tab_coeffs} we summarize our results for
the calculated model parameters depending on the weight choice.
As expected, all parameters (with the exception of the 
elastic coefficient) considerably differ between
the equal-mass and the standard-mass convention.
Recall that this difference is twofold. First,
there is a trivial scale factor (proportional to
some power of $P_0$) that is due to 
the normalization of the polar eigendisplacements 
$p_\kappa$.
The present choice of measuring the polar order 
parameter via the norm (in length units) of the atomic distortion 
amplitude differs from the usual convention of macroscopic 
theories. To connect with the latter we provide in Table~\ref{tab_si}
\black{
(Appendix C)
}
the same coefficients in SI units, where we have set $P_0$ 
to the spontaneous polarization of the bulk crystal, in C/m$^2$.
In this case, the $A$ and $B$ coefficients agree between different
choices of the weights, while a discrepancy remains in both $f$ 
and $G$.
Such a dependence of $f$ and $G$ on the weight choice relates
to the ambiguity in the definition of the center of mass, which 
is subtracted by construction from $p_\kappa$ via Eq.~(\ref{conda}).
As we said, either source of arbitrariness must have no impact 
on the physical results that we extract from the model --
we shall use this criterion to validate the internal 
consistency of our theory.

As a first test of such claim, we use the calculated values of the 
coefficients to 
%Armed with the values of the coefficients for our materials, we can now
%use the Landau model to 
compute the energy and the width of the domain walls
following Equations \ref{eqxi} and \ref{eqdwe}.
We find that the results are indeed consistent (to machine precision)
between the two weight choices.
Table \ref{tab_prop} shows that the domain-wall energies extracted from the
model are also consistent with the results of our full DFT calculations; the
level of agreement is remarkable considering the simplicity of the Landau model.
The wall width, in particular, is of the order of the cubic lattice parameter in 
all cases, which provides a rather difficult test for the continuum approximation.
(The latter is expected to break down at length scales that are comparable with the
lattice periodicity.)
Note that the direct DFT calculations yield a marked dependence of the
domain wall energy on the wall location respect to the underlying atomic 
structure; such a dependence is obviously missing within the continuum 
description.

%\black{Next, we calculate the atomic distortions
%}

\begin{figure}%[h]
\centering
\includegraphics[width=70mm,angle=0]{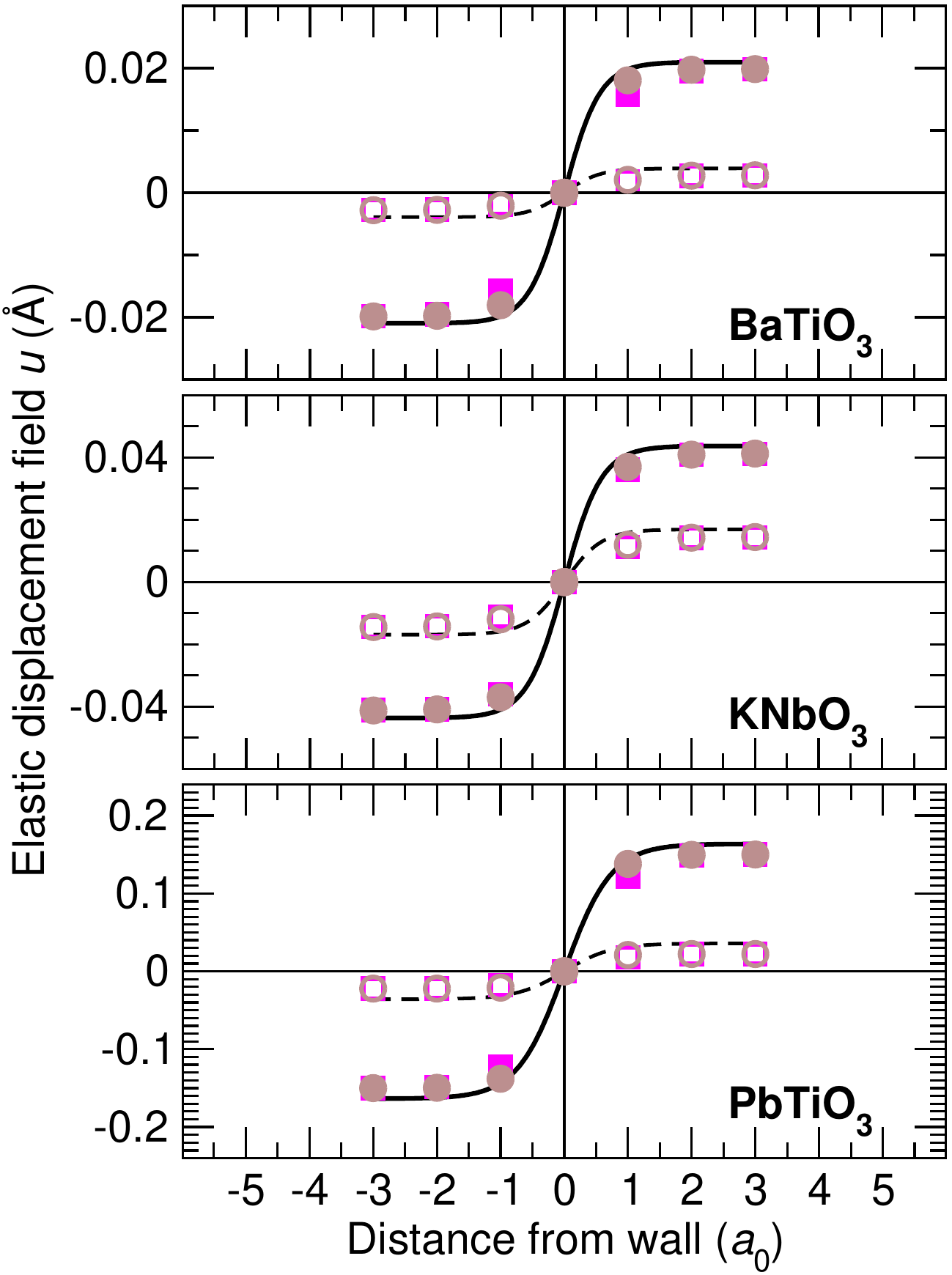}
\caption{(Color online.)
Value of the elastic displacement field as a function of distance to the 
domain wall, for three perovskite oxides.
The lines correspond to Landau model predictions using unit weights (broken
lines) or atomic weights (continuous lines).
The symbols correspond to local center of mass calculations computed with
full DFT, for different cells centered on $A$O or $B$O$_2$ layers, and for
those two different weights.
}
\label{fig_u}
\end{figure}

Next, we compute the elastic displacement field by combining
Eqs.~(\ref{eq_u}) and ~(\ref{eqtanh}),
\begin{equation}
u(x) = -\frac{f}{C}P(x) = -\frac{f}{C} P_0 \tanh \frac{x}{\xi}.
\end{equation}
The resulting profile, shown in Fig.\ \ref{fig_u}
drastically differs depending on the weight choice,
but is in excellent agreement with the 
``local center of mass'' that we extract from our 
explicit domain-wall calculations by using the same weight 
convention. (For an $A$O wall, we compute this by
considering the positions of the $A$ and O atoms in a $A$O layer together
with an average of the $B$ and O atoms in adjacent $B$O$_2$ layers---we
follow an equivalent procedure for the $B$O$_2$ walls.)
%
%which will have a different profile depending on the weights chosen, as seen
%in the curves of Fig.\ \ref{fig_dwpattern}.
In all materials considered here, taking unit weights 
(discontinuous lines) leads to flatter profiles than taking atomic
weights (continuous lines), but with neither choice the centers 
of mass is fully aligned across the wall.
Any choice of weights, on the other hand, leads to the same values of atomic
displacements predicted by the Landau model via Eq.~(\ref{basis}).
%\begin{equation}
%u_\kappa(x) = \left(-\frac{f}{C}+p_\kappa \right) P_0 \tanh \frac{x}{\xi};
%\end{equation}
%Table \ref{tab_prop} lists the displacements of the $A$ ions, $p_A$, showing
%remarkable agreement between the values computed with DFT and the ones
%of the Landau model.
The resulting curves are plotted as continuous lines next
to the full DFT results for the domain-wall structures in 
% these model results 
%In the case of BaTiO$_3$ 
%we have plotted a more detailed comparison in 
Fig.\ \ref{fig_dwpattern}. 
The agreement between the displacements
predicted by the Landau model and those from 
full DFT calculations is obvious.

For a more quantitative analysis, we extract the overall 
displacement of the center of mass across the wall, $\Delta u$, from the relaxed  
domain-wall structures, and use it to estimate the effective
flexocoupling coefficient by inverting Eq.~(\ref{offset}),
\begin{equation}
\label{feff}
f^{\rm eff} = - C\frac{\Delta u}{2P_0}.
\end{equation}
%compare it with the prediction of the macroscopic
%model; 
The results for both choices of weights are compared in Table \ref{tab_feff}
to the values predicted by the macroscopic model.
One can note that the equal-weight convention yields values of 
$f$ and $f^{\rm eff}$ that are systematically smaller than those
obtained by setting $w_\kappa$ to the physical atomic masses.
Overall, the values of $f^{\rm eff}$ and $f$ nicely agree, differing
at most by few tenths of a volt in all cases.

Interestingly, the continuum approach yields a slight
overestimation of the converse flexoelectric effect at the wall;
such a feature appears to be systematic across all the materials set and 
irrespective of the weight convention being used.
A possible explanation might lie in the atomically sharp nature
of the domain-wall structures, which clearly challenges the
continuum description.
The estimated values of $f^{\rm eff}$, however, appear to be largely 
insensitive to the atomistic details of the wall (AO- and BO$_2$-type 
walls yield very similar values for most materials); therefore, it is 
unlikely that the aforementioned discrepancy originate from the continuum 
approximation itself.
We suspect that such an effect may depend on higher-order terms (either
in the gradient expansion or in the polar distortion amplitude) that 
we neglect in Eq.~(\ref{free}).
In any case, the accuracy of the present theory is more than sufficient
for a quantitative comparison of first-principles calculations and 
experiments, e.g., along the lines of Ref.~\onlinecite{Wang-20}.

%\begin{table}
%\begin{center}
%\begin{tabular}{rddd}
%\hline \hline
%& \multicolumn{1}{c}{DFT ($A$O)}
%& \multicolumn{1}{c}{DFT ($B$O$_2$)}
%& \multicolumn{1}{c}{Landau Model} \\
%\hline
%BaTiO$_3$ &  0.00281 & 0.00279 & 0.00391 \\
%CaTiO$_3$ &  0.0204  & 0.0226  & 0.0339 \\
% KNbO$_3$ &  0.0145  & 0.0144  & 0.0169 \\
%NaNbO$_3$ &  0.0511  & 0.0515  & 0.0634 \\
%PbTiO$_3$ &  0.0226  & 0.0228  & 0.0358 \\
%PbZrO$_3$ & -0.00403 & 0.0128  & 0.0392 \\
%\hline
%BaTiO$_3$ &  0.0199  & 0.0198  & 0.0210 \\
%CaTiO$_3$ &  0.0761  & 0.0783  & 0.0896 \\
% KNbO$_3$ &  0.0412  & 0.0412  & 0.0437 \\
%NaNbO$_3$ &  0.0834  & 0.0838  & 0.0957 \\
%PbTiO$_3$ &  0.150   & 0.151   & 0.164 \\
%PbZrO$_3$ &  0.214   & 0.231   & 0.258 \\
%\hline \hline
%\end{tabular}
%\caption{
%Displacement of the center of mass of a cell in the middle of a domain, 
%$\Delta u$, as computed with DFT and with our Landau model.
%Top block: unit masses are used. Bottom block: actual
%atomic masses are used.
%}
%\label{tab_coeffs}
%\end{center}
%\end{table}

\begin{table}
\begin{center}
\begin{tabular}{cddd}
\hline \hline
& \multicolumn{1}{c}{DFT ($A$O)}
& \multicolumn{1}{c}{DFT ($B$O$_2$)}
& \multicolumn{1}{c}{Landau Model} \\
\hline
BaTiO$_3$     & -0.203 & -0.198 & -0.283 \\
BaTiO$_3^{(*)}$ & -0.279 & -0.279 & -0.288 \\
CaTiO$_3$     & -0.409 & -0.454 & -0.681 \\
 KNbO$_3$     & -0.676 & -0.670 & -0.793 \\
NaNbO$_3$     & -1.07  & -1.09  & -1.33  \\
PbTiO$_3$     & -0.395 & -0.400 & -0.628 \\
PbZrO$_3$     &  0.038 & -0.124 & -0.373 \\
\hline
BaTiO$_3$     & -1.43  & -1.43  & -1.52  \\
BaTiO$_3$$^*$ & -1.47  & -1.48  & -1.49  \\
CaTiO$_3$     & -1.53  & -1.57  & -1.80  \\
 KNbO$_3$     & -1.93  & -1.92  & -2.05  \\
NaNbO$_3$     & -1.75  & -1.76  & -2.01  \\
PbTiO$_3$     & -2.63  & -2.64  & -2.87  \\
PbZrO$_3$     & -2.04  & -2.20  & -2.45  \\
\hline \hline
\end{tabular}
\caption{
Value of $f^{\rm eff} = -C \Delta u / 2 P_0$, in V, 
as computed with DFT and with our Landau model.
Top block: unit masses are used. Bottom block: actual
atomic masses are used.
}
\label{tab_feff}
\end{center}
\end{table}

\subsection{``Soft'' ferroelectric walls: BaTiO$_3$ under pressure}

\label{sec:pressure}

The values of $\xi$ in Table \ref{tab_prop} are smaller than one lattice
parameter in all cases, which
agrees with the usual perception that 180$^\circ$ ferroelectric domain
walls are very thin.
In order to evaluate the accuracy of the polarization profile of 
Equation \ref{eqtanh} it would be desirable to study thicker ferroelectric 
domain walls with a smaller distortion amplitude; in such a limit we expect 
the free energy functional of Eq.~(\ref{free}) to match the results of
direct DFT calculations \emph{exactly}.
One way to access this regime consists in using an external parameter 
to bring the material closer to the phase transition. 
%e.g.,  
%by working at finite temperature. 
%
[The domain wall thickness, also known as \emph{correlation length},
diverges in a vicinity of a ferroelectric phase transition, because of the
vanishing $A$ coefficient at the denominator of Eq.~(\ref{eqxi}).]
Temperature is the most obvious choice in an experimental context; this is, 
however, impractical in the context of direct DFT simulations.
A simpler alternative consists in applying a hydrostatic pressure $p$ to our simulation
cells; in many ferroelectrics, this results in a suppression of the ferroelectric instability
already at moderate values of $p$, thus mimicking the effect of increasing temperature in real
experiments. 
%as the soft mode eigenvalue approaches zero the atomic displacements
%around the wall become smoother.

\begin{figure}%[!t]
\centering
\includegraphics[width=80mm,angle=0]{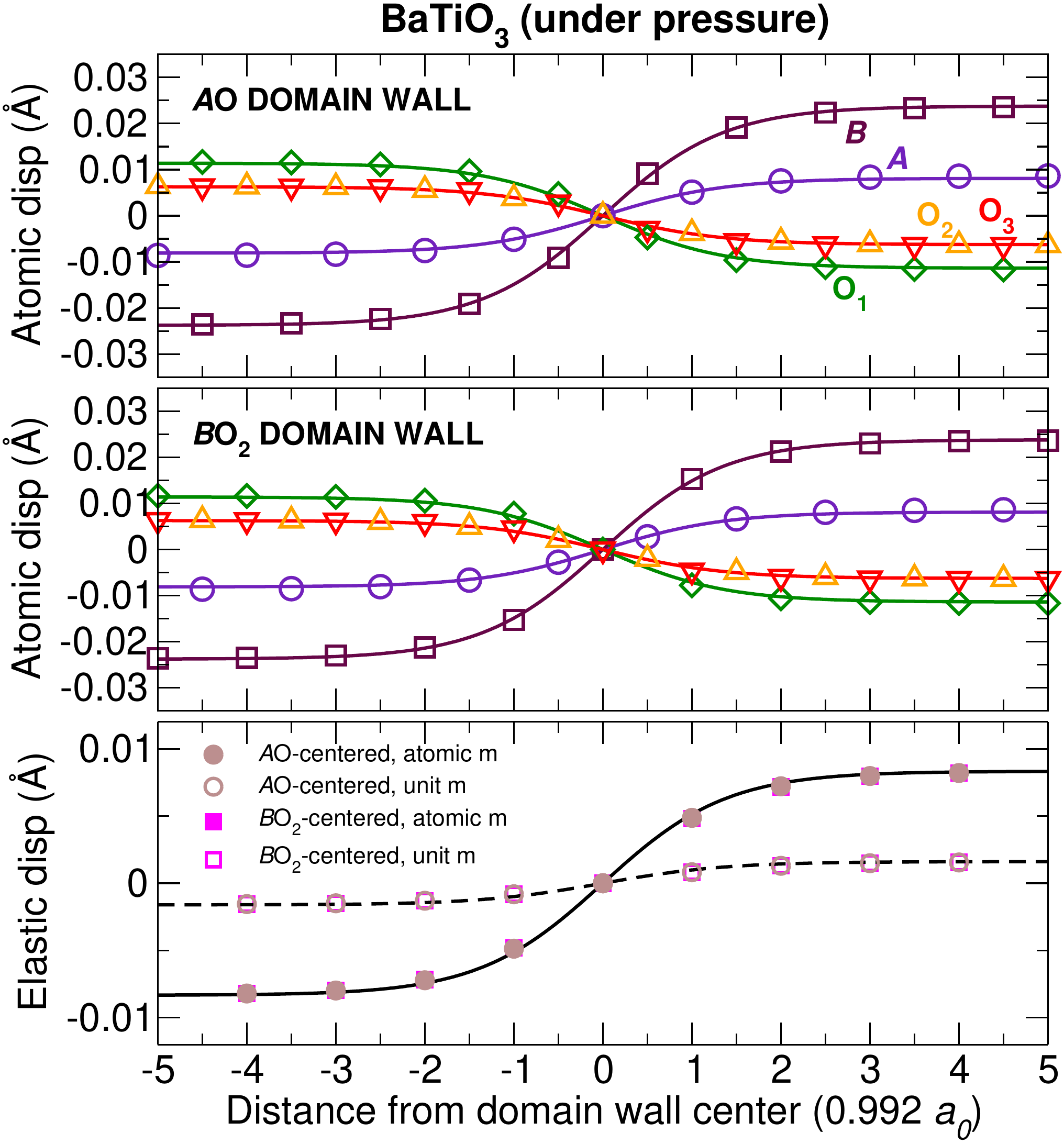}
\caption{(Color online.)
Atomic displacements parallel to an $A$O domain wall (top panel)
and to a $B$O$_2$ wall (middle panel), and elastic displacement (bottom
panel), as a function of the distance from the domain wall.
In all cases the system studied is compressed BaTiO$_3$, built
using supercells with lattice a parameter of $0.992 a_0$.
}
\label{fig_dispBTO}
\end{figure}

We shall apply this strategy to BaTiO$_3$ by repeating all the calculations
under hydrostatic pressure; the corresponding results are indicated in 
the tables with a (*) symbol. 
We find that, at a lattice constant of 99.2\% of the equilibrium $a_0$
(see Table \ref{tab_fiveatomcells}), the $A$ 
parameter is still negative, but its \black{absolute value} is about an order of 
magnitude smaller (see Table~\ref{tab_coeffs}) than in standard conditions.
Consistently, as the soft mode eigenvalue approaches zero the atomic displacements
around the wall become smoother, and the
value of $\xi$ increases to 5.441~\AA.
The profile of atomic displacements for these walls is shown in Figure
\ref{fig_dispBTO}: the results of the Landau model (continuous
lines, parameters are reported in Table~\ref{tab_coeffs}) 
are essentially in perfect agreement with the full DFT results in a
$20 a_0 \times a_0 \times a_0$ unit cell (symbols),
both for the $A$O and for the $B$O$_2$ types of domain wall.
The agreement regarding domain wall energy is also extremely good (see Table~\ref{tab_prop}): 
Equation (\ref{eqdwe}) yields 0.324 mJ/m$^2$, while from full DFT calculations we obtain
0.321 mJ/m$^2$ (for the $A$O type) and 0.322 mJ/m$^2$ (for the $B$O$_2$ type).
% (Further numerical details about the calculated parameters %that led to these results 
% are given in the Appendix.)

Note that the direct DFT calculations of the domain wall energies 
and atomic structures are numerically much more challenging than in the 
zero pressure case, because of the extreme softness of the ferroelectric instability.
To achieve a reasonable level of accuracy, and hence allow for a meaningful
comparison to the model results, we had to be unusually careful with 
the usual computational parameters: (i) the domain-wall calculations 
were performed with a larger $20 a_0 \times a_0 \times a_0$ unit cell,
to accomodate the thicker wall structure; (ii) the five-atom unit cell 
calculations were done using a $20 \times 8 \times 8$ Monhkhorst-Pack grid, 
exactly reproducing the folded Brillouin zone of the supercell (the 
ferroelectric distortion amplitude and double-well potential depth is 
remarkably sensitive to the ${\bf k}$-point mesh); (iii) the tolerance on 
residual forces were set to 0.0001 eV/~\AA,
approaching the inherent precision limits of the numerical algorithms.
On the other hand, the calculation of the Landau model parameters required a
similar computational effort as in the zero-pressure case, highlighting the 
obvious advantages of our multiscale approach in softer ferroic materials.
%was exactly as 
%expensive as in the 
%It is important in this situation to control possible sources of numerical noise
%to allow for a meaningful comparison of the Landau model and the full DFT
%calculation.
%To achieve this, since the domain wall DFT results
%were obtained in a $20 a_0 \times a_0 \times a_0$ unit cell.

%%%%%%%%%%%%%%%%%%%%%%%%%%%%%%%%%%%%%%%%%%%%%%%%%%%%%%%%%%%%%%%%%%%%%%%%%%%%%%%%

\section{Discussion}

\label{sec:discussion}

The impact of flexoelectricity on the properties of ferroelectric
domain walls was studied in several recent 
works.~\cite{Yudin-12,Yudin-13,Gu-14,Wang-19,Wang-20}
Yudin and Tagantsev~\cite{Yudin-13} established the role of
flexoelectricity in determining the elastic offset at the wall,
as well as its impact on domain wall energy and thickness 
via the renormalization of the polarization gradient coefficient, $G$.
These formal results, which we have largely built upon in our
present work, were applied by
Wang {\em et al.}~\cite{Wang-20} to domain-wall structures 
in PbTiO$_3$ that were obtained either via first-principles
calculations or experimental microscopy images.
These authors, however, assumed that an ``authentic''~\cite{Wang-20}
definition of the strain exists, and overlooked (as did earlier
works on this topic) the inherent arbitrariness that
we highlight here.
The definition used by Ref.~\onlinecite{Wang-20} 
corresponds to using physical masses as weights within our
formalism. Within such a convention, however, our result of
$f_{44}=-$2.87 V (or $f^{\rm eff}_{44}=-$2.64 V as extracted
from the direct calculation of the BO$_2$-centered domain-wall 
structure) in PbTiO$_3$ disagrees  
with the value of $f_{44}=-$5.4 V quoted by Wang {\em et al.}
by a factor of two.
We believe that the source of the disagreement lies in 
Eq.~(9b) of Wang {\em et al.}~\cite{Wang-20}, where an additional
factor of two is indeed present compared to our Eq.~(\ref{feff}).
%
%Such a factor of two is likely wrong in Ref.~\onlinecite{Wang-20}: 
Note that our calculated $C=101$ GPa is in excellent agreement 
with the value of $C_{44}=100.8$ GPa quoted therein, ruling out 
possible issues in the definition of the shear elastic constant.
Once the result of Ref.~\onlinecite{Wang-20} is divided
by two, it is in excellent agreement with ours.
%
%Other authors,~\cite{Yudin-12,Gu-14} focused on secondary
%components of the polarization that would emerge at ferroelectric 
%walls due to flexoelectricity, leading to bichiral structures.
%
%While the study of these additional features goes beyond the 
%scopes of our work, we can anticipate that the principle 
%of invariance established here also applies to the more
%complex free-energy functionals of Refs.~\cite{Yudin-12,Gu-14};
%a detailed account of this topic will be given in a forthcoming 
%publication.

From the point of view of the physics, the main conceptual advance
of our work can be summarized as follows:
The contribution of flexoelectricity to the domain-wall properties
is ill-defined; however, the contribution of the polarization gradient
energy is ill-defined as well, and including both terms is essential for 
for guaranteeing that their respective arbitrariness cancels out. 
The obvious question then is: Is it really necessary to consider
both terms explicitly? Or can we choose the weights in such a way that
the contribution of flexoelectricity vanishes identically, leaving only
the gradient terms?
A quick glance at Fig.~\ref{fig_u} suggests, at first sight, 
that the answer to the latter question be ``yes'':
for the equal-weights case, the elastic offset between 
the domains is already small -- by weighing the cations
slightly less than the oxygens, one could certainly 
make $\Delta u$ to vanish exactly.
However, this would ``renormalize out'' flexoelectricity 
only for a [100]-oriented wall; the same choice 
of weights would not yield a vanishing $\Delta u$ 
for a different (say, [110]) orientation.
This means that, for a truly isotropic material, 
flexoelectricity can be reabsorbed into the polarization
gradient energy, while anisotropic crystals generally
require its explicit treatment.
\black{%Another way to express this idea is to 
In other words, one can equivalently state that
the renormalized gradient coefficients, $\tilde{G}$, contain 
a \emph{nonanalytic elastic contribution},
which generally prevents their representation in a straightworward
tensorial form.
This adds up to the more conventional sources of nonanaliticity,
due to the long-range electrostatic interactions, which we have
not covered in the present work.
Interestingly, electrostatic and elastic interactions mediated 
by flexoelectricity share many similarities, as the former also enjoy a 
covariance principle~\cite{Stengel-16} and become analytic 
(i.e., short ranged) in isotropic media.~\cite{Stengel-16}} 
%providing an interesting parallel 

%Similar conclusions were reached in Ref.~\onlinecite{Stengel-16}
%via a consideration of the electrostatic energy in presence
%of an inhomogeneous deformation of the crystal.
%
\black{Another conceptual advance of this work consists in establishing a 
first-principles theory of the polarization gradient coefficient, $G$.
The established procedure to infer the value of $G$ from the 
phenomenological point of view consists in analyzing the dispersion
of the phonon band associated with the soft mode.~\cite{Hlinka-06}
The conceptual difficulties of defining $G$ in a microscopic context,
and the often counterintuitive consequences of combining LGD models 
with density-functional theory, have been emphasized very recently.~\cite{Samanta-22}
(For example, an estimation based on the calculated 
spontaneous polarization and domain-wall energy resulted~\cite{Samanta-22}
in a large variation of $G$ as a function of the applied pressure.) 
%motivated the authors to abandon LGD theory in favor of a lattice model.
%
In contrast with the results of Ref.~\cite{Samanta-22}, the $G$ coefficient
as defined in our work changes little (2--3\% deviation for an isotropic 
compressive strain of 1\%) with external pressure in BaTiO$_3$, 
consistent with the usual assumptions of LGD theory.
This, together with the excellent accuracy of the domain wall energies
and structures calculated within our continuum approach, 
allows us to reach a more optimistic conclusion (compared to Ref.~\cite{Samanta-22})
regarding the suitability of LGD equations for the description of realistic 
domain-wall structures.
%
% shows a wild 
%variation with pressure 
%Our approach allows to settle many of these issues
%and result in a closer synergy between continuum-based and first-principles
%studies.
%We believe that the present approach resolves many of these issues:
%For example, 
% and 
%
%
}

Before closing, it is useful to place the methodology
that we have developed here in the context of the existing literature.
Our strategy differs in spirit from Landau-Ginzburg-Devonshire (LGD)
theories in that the latter have been traditionally approached
with a phenomenological mindset.
Granted, combining first-principles techniques with LGD theories
is not new: the former are being increasingly used~\cite{Artyukhin-14} 
to estimate a subset, or even the entirety of the LGD model coefficients.
% that might prove difficult
%to infer from the experiments alone.
%
However, such a practice is seldom supported by a formal link
between the macroscopic and microscopic degrees of freedom, 
which has thwarted so far a quantitative validation of
LGD models against the \emph{ab initio} results.
(At the qualitative level, successful comparisons do exist,
see Ref.~\onlinecite{Gu-14} for an example that is relevant to 
the topics of this work.)
The conceptual novelty of our approach consists in deriving the
macroscopic field equations as a rigorous long-wave approximation
to the discrete lattice Hamiltonian.
This constitutes a much more intimate synergy between the two
levels of theory, which allows for quantitatively accurate
predictions (compared to the ``training'' first-principles model)
of the relevant physical properties.
Most importantly, our approach leads to a
deeper awareness of the internal structure of the theory,
the mutual relation between the many physical quantities
involved, and their potential dependence on some arbitrary
choices that are inevitable in the mapping of the problem 
onto continuum fields.

For the above reasons, our ``first-principles macroscopic
theory'' belongs to the class of 
methods that are commonly known as \emph{second-principles}.
The latter are obtained via an appropriate coarse-graining of the
first-principles Hamiltonian, whose physically relevant low-energy
degrees of freedom are treated explicitly, while most of the
original complexity is integrated out.
In the case of ferroelectrics, the reference in this context
is the ``effective Hamiltonian'' method, pionereed by
Zhong, Rabe and Vanderbilt;~\cite{Zhong-95} \black{many 
improvements and generalizations thereof have been 
introduced over the years.}~\cite{Wojdel-13,Ghosez-22}
%Ghosez-22: https://www.annualreviews.org/doi/10.1146/annurev-conmatphys-040220-045528
%authors have
%approached the problem in a similar spirit over the years.
%
Performing a fair comparison of the virtues and shortcomings 
of either strategy would require substantial additional work,
and will be best carried out in the framework of a separate
publication.
Here we will limit ourselves to observe that, once the
continuum differential equations are discretized on a
regular mesh corresponding to the unit cells of the 
original perovskite lattice, our approach essentially 
reduces to an effective Hamiltonian -- at least for 
the simple domain-wall geometry considered here.
%on a finite-elements grid, 
%
[A generalization to the full three-dimensional case
appears feasible, too, by writing all the couplings of 
Eq.~(\ref{free}) in a tensorial form and by explicitly 
treating the electrostatic energy.]
From this point of view, the present method effectively
bridges the gap between atomistic and continuum approaches,
%Second-principles methods (atomistic) vs. phenomenological (continuum):
%our approach bridges between the two.
while preserving an exact limit at length scales that 
are large compared to the interatomic spacings.

\section{Conclusions}

\label{sec:conclusions}

We have established a formal mapping between continuum fields and atomistics,
and demonstrated its predictive power in the study of spatially inhomogeneous
structures (e.g. domain walls) in ferroics.
Our formalism demonstrates the necessity of abandoning some
widespread beliefs in continuum theory, for example that
the local strain field be a 
physically well-defined degree of freedom of the crystal.
While the arbitrariness in the definition of the elastic strain 
has a profound impact on the continuum model coefficients,
we demonstrate that the physical answers derived from
the model are robust against the specific convention that
is being used.

On one hand, this results in a fundamental \emph{principle of 
invariance} that we deem of great practical utility in validating
the internal consistency of the continuum equations.
On the other hand, and most importantly, our results 
provide a stringent benchmark to determine what are 
the physically sound questions that one can ask, and
what are not.
As illustrated by our practical tests, examples of well-posed questions 
concern the domain wall energy, or the atomic positions; 
we show that macroscopic theory (within the validity range
of the continuum approximation) can be an excellent tool to predict both --
with comparable accuracy to the full first-principles ``training model''.
Conversely, questions of the type 
``what is the contribution of flexoelectricity to the domain wall energy?'', or
``what is the impact of the dynamic flexoelectric effect on the 
acoustic phonon dispersion?''
are physically meaningless, as the answer can be about ``anything'', 
depending on some (necessarily) arbitrary choices that one makes along the way.
% any theory trying to answer 
%them is doomed to failure.

In the present work we have voluntarily chosen, for the sake of clarity, 
a minimal model of macroscopic phenomena where the above ideas have a nontrivial 
impact. (For example, we have neglected most components of the strain 
tensor, as well as competing antiferrodistortive modes.)
This choice has inevitably limited the predictive power
of our study in some materials.
An obvious future development of this work consists in extending the 
scopes of our first-principles continuum approach to more complex 
structures, e.g., involving a higher dimensionality or a broader range of
degrees of freedom. 
%
%Other authors,~\cite{Yudin-12,Gu-14} focused on secondary
%components of the polarization that would emerge at ferroelectric 
%walls due to flexoelectricity, leading to bichiral structures.
%
%While the study of these additional features goes beyond the 
%scopes of our work, we can anticipate that 
For example, it will be interesting to clarify whether the principle 
of invariance established here also applies to the bichiral
domain-wall structures described in Refs.~\cite{Yudin-12,Gu-14},
\black{or to the secondary Bloch-like components that were
theoretically predicted in PbTiO$_3$~\cite{Wojdel-14}}.
%Wojdel-14: https://journals.aps.org/prl/abstract/10.1103/PhysRevLett.112.247603
%https://iopscience.iop.org/article/10.1088/0953-8984/25/30/305401
%
%a detailed account of this topic will be given in a forthcoming 
%publication.
Also, incorporating the effect of 
octahedral tilts appears especially promising, in
light of the results of Ref.~\cite{Schiaffino-17}.
Generalizing the ideas developed here in such directions will be an exciting
avenue for further study.

%is unavoidable in any macroscopic theory that combines 
%The strongest philosophical resistance against our arguments
%consists in the widespread belief that the local strain field be a 
%physically well-defined degree of freedom of the crystal.
%
%How can a theory yield arbitrary answers for a quantity that
%can be deterministically extracted from the relaxed atomic positions?
%
%Such a theory \emph{must} be wrong -- unless there were some 
%hidden subtleties in the ``deterministic'' procedure that 
%one uses to perform the inverse operation of Eq.~(\ref{micro}),
%i.e., the mapping from atomistic to continuum.

%%%%%%%%%%%%%%%%%%%%%%%%%%%%%%%%%%%%%%%%%%%%%%%%%%%%%%%%%%%%%%%%%%%%%%%%%%%%%%%%

\acknowledgements

O.D. acknowledges funding from the Israel Science Foundation under Grants No.
1814/14 and No. 2143/14.
 M.S. acknowledges the support of Ministerio de Economia,
 Industria y Competitividad (MINECO-Spain) through
 Grant No. PID2019-108573GB-C22 and Severo Ochoa FUNFUTURE 
 center of excellence (CEX2019-000917-S);
 of Generalitat de Catalunya (Grant No. 2017 SGR1506);
 and of the European Research Council (ERC) under the European Union's
 Horizon 2020 research and innovation program (Grant
 Agreement No. 724529). 

%%%%%%%%%%%%%%%%%%%%%%%%%%%%%%%%%%%%%%%%%%%%%%%%%%%%%%%%%%%%%%%%%%%%%%%%%%%%%%%%

\appendix

\section{Relation to the eigenmode representation}
\label{sec_eigrep}

In this Appendix we shall %use the above considerations to generalize
link the formalism developed in the main text to
the prescriptions of Ref.~\onlinecite{Stengel-16} for the
construction of the continuum Hamiltonian. 
Their proposed strategy consists in identifying the distortion patterns associated 
with the mechanical displacement and polarization fields with, respectively, 
the acoustic and ``soft'' polar eigenmodes of the zone-center dynamical matrix, $\mathcal{D}$.
We shall adopt here a slightly more general definition by introducing the
following operator,
\begin{equation}
\widetilde{\mathcal{D}}_{\kappa \alpha, \kappa' \beta} = \frac{1}{\sqrt{w_\kappa w_{\kappa'}}} \Phi^{(0)}_{\kappa \alpha, \kappa' \beta},
\end{equation}
which reduces to the standard definition of $\mathcal{D}$ when the 
weights are set to the physical atomic masses, $m_\kappa$.
\black{(In our formalism the weights are dimensionless, so the appropriate
choice in this case consists in setting them to the \emph{fractional}
mass of the sublattice; then, $\widetilde{\mathcal{D}}$ reduces to
$M_{\rm tot} \mathcal{D}$, where $M_{\rm tot}=\sum_\kappa m_\kappa$ is
the total mass of the cell.)
Following Ref.~\onlinecite{Stengel-16}, we define the eigendisplacement pattern $|\nu \alpha \rangle$
associated to a given normal mode $\nu \alpha$ ($\nu$ runs over the four $T_{1u}$ irreps of the cubic
perovskite structure, including acoustic and optical TO$_1$--TO$_3$ modes; $\alpha$ is a Cartesian 
direction) as
\begin{equation}
\langle \kappa \beta|\nu \alpha \rangle = \sqrt{\frac{1}{w_\kappa}}   \langle \kappa \beta|v^{(\nu)}_\alpha \rangle,
\end{equation}
where $|v^{(\nu)}_\alpha \rangle$ are the normalized eigenvectors of $\widetilde{\mathcal{D}}$.
(The present use of the bra-ket notation follows the conventions established in earlier works:~\cite{Stengel-16,Zabalo-21}.
Both bras and kets are real vectors in a space of dimension $3N_{\rm at}$, where $N_{\rm at}$ is the total number of atoms
in the unit cell; $\langle \kappa \beta|$ form a complete orthonormal basis.)}
This definition results in a generalized orthonormality condition for the eigendisplacements,
\begin{equation}
\label{ortho}
\langle \nu \alpha | W  | \nu' \beta \rangle = \delta_{\nu \nu'} \delta_{\alpha \beta},
\end{equation}
where the ``overlap operator'' $W$ is defined as
\begin{equation}
W_{\kappa \alpha,\kappa'\beta} = w_\kappa %\frac{ w_\kappa} { \sum_\kappa w_\kappa } 
\delta_{\kappa \kappa'} \delta_{\alpha \beta}.
\end{equation}
One can show~\cite{Stengel-16} that, within such prescriptions, the eigendisplacements 
associated to the acoustic mode ($\nu=u$) are independent of the weights, thus recovering Eq.~(\ref{ub}).
Then, by setting $\nu=u$ and $\nu'=$TO$_1$--TO$_3$ in Eq.~(\ref{ortho}), we find that 
both Eq.~(\ref{conda}) and Eq.~(\ref{condb}) are automatically satisfied by all polar modes. 
\black{We can, thus, identify the symbol
$\langle \kappa \beta|\nu \alpha \rangle$ with the transformation matrix introduced in Section~\ref{sec:direct},
\begin{equation}
\langle \kappa \beta|\nu \alpha \rangle = T_{\kappa \beta,\nu \alpha}.
\end{equation}
In particular, the elastic displacement and polarization basis vectors are defined by
\begin{equation}
\label{evec}
\langle \kappa \beta|u \alpha\rangle = \delta_{\alpha\beta}, \qquad \langle \kappa \beta|{\rm TO}_1  \alpha\rangle = 
\delta_{\alpha\beta} p^{(\alpha)}_{\kappa},
\end{equation}}
where we have assumed that TO$_1$ corresponds to the ferroelectric ``soft mode''.

%\subsection{The polar eigendisplacement vector}

\begin{table*}%[!b]
\begin{center}
\begin{tabular}{cdddddd}
\hline \hline
& \multicolumn{1}{c}{$A$ ($10^9$m/F)}
& \multicolumn{1}{c}{$B$ ($10^{10}$$\Omega^2$m$^3$/kg)}
& \multicolumn{1}{c}{$C$ ($10^{11}$Pa)}
& \multicolumn{1}{c}{$f$ (V)}
& \multicolumn{1}{c}{$G$ ($10^{-10}$m$^3$/F)} 
& \multicolumn{1}{c}{$\tilde{G}$ ($10^{-10}$m$^3$/F)} \\
\hline
BaTiO$_3$     & -0.857 & 2.43  & 1.36  & -0.283 & 0.214 & 0.208  \\
BaTiO$_3^{(*)}$ & -0.145 & 2.36  & 1.40  & -0.288 & 0.221 & 0.215  \\
CaTiO$_3$     & -1.66  & 0.645 & 1.02  & -0.681 & 0.550 & 0.505  \\
 KNbO$_3$     & -0.934 & 2.22  & 0.961 & -0.793 & 0.313 & 0.248  \\
NaNbO$_3$     & -1.56  & 1.21  & 0.755 & -1.33  & 0.746 & 0.510  \\
PbTiO$_3$     & -1.00  & 0.299 & 1.01  & -0.628 & 0.447 & 0.409  \\
PbZrO$_3$     & -4.15  & 0.991 & 0.616 & -0.373 & 1.00  & 0.979  \\
\hline
BaTiO$_3$     & -0.857 & 2.43  & 1.36  & -1.52  & 0.378 & 0.208  \\
BaTiO$_3^{(*)}$ & -0.145 & 2.36  & 1.40  & -1.49  & 0.373 & 0.215  \\
CaTiO$_3$     & -1.66  & 0.645 & 1.02  & -1.80  & 0.822 & 0.505  \\
 KNbO$_3$     & -0.934 & 2.22  & 0.961 & -2.05  & 0.684 & 0.248  \\
NaNbO$_3$     & -1.56  & 1.21  & 0.755 & -2.01  & 1.05  & 0.510  \\
PbTiO$_3$     & -1.00  & 0.299 & 1.01  & -2.87  & 1.22  & 0.409  \\
PbZrO$_3$     & -4.15  & 0.991 & 0.616 & -2.45  & 1.95  & 0.979  
\\
\hline \hline
\end{tabular}
\caption{
Coefficients (in SI units) of the Landau model of Eq.~(\ref{free})
computed from DFT calculations, when $P_0$ is the spontaneous 
polarization of the bulk crystal (measured in C/m$^2$).
Top block: unit weights are used. Bottom block: physical
atomic masses are used.
}
\label{tab_si}
\end{center}
\end{table*}

There is a slight drawback with such a procedure: different conventions for 
the weights lead to definitions of $p^{(\alpha)}_{\kappa}$ that are generally not 
related via Eq.~(\ref{ppr}). 
%
%First, there is a trivial scale factor, related to the normalization condition on
%the eigenvectors of $\widetilde{\mathcal{D}}$; this is, however, irrelevant as 
%it only affects the length unit that we use to ``measure'' the polar distortion,
%which is to a large extent arbitrary. 
%
%Second, and most importantly
Indeed, the configuration space spanned by $p^{(\alpha)}_{\kappa}$ 
changes depending on the weights, as the three polar optical modes of the perovskite 
structure can mix. 
(This is a well-known issue in the construction of effective Hamiltonian for 
ferroelectrics, where typically the lowest eigenvector of the force-constant matrix
is used for $p^{(\alpha)}_{\kappa}$; this corresponds to choosing equal weights in 
the context of our formalism.)
In practical cases, $p^{(\alpha)}_{\kappa}$ might not reproduce the correct distortion pattern 
(and energetics) of the bulk ferroelectric ground state, which is undesirable in the
study of a domain wall. 
%(The description of the domains needs to be as accurate as
%possible.)
%
To avoid this issue, in this work we have followed the prescriptions of 
Section ~\ref{sec:converse} and defined $p^{(\alpha)}_\kappa$ starting from the relaxed 
distortion pattern of the bulk ferroelectric crystal instead.
%and use the following procedure to define the
%polar eigendisplacements $p_\kappa$.
%
%orthonormalize it to the acoustic mode
%according to Eq.~(\ref{ortho}). This uniquely defines $|P_\alpha\rangle$ 
%for a given choice of $w_\kappa$, and is consistent with the results 
%of the previous Section.
%
In the limit of a weak ferroelectric instability, one can show that 
this definition of $p^{(\alpha)}_{\kappa}$ exactly matches the eigenvector 
representation provided by Eq.~(\ref{evec}) regardless of the choice of the 
weights.

\section{Dynamical equations of motion}

\label{sec_dyn}

Tagantsev~\cite{tagantsev-86} and Kvasov and Tagantsev~\cite{Kvasov-15} claimed 
that there are two well-defined contribution to the bulk flexoelectric tensor, static 
and dynamic in nature, and that they are, in principle, separately measurable. 
Later works~\cite{artlin} clarified that such a partition is arbitrary,
and that: (i) the total flexoelectric coefficient is meaningful for
dynamical problems; (ii) either the total or the ``static'' flexoelectric
tensor yield identical answers at mechanical equilibrium.
To firm up our arguments, we shall revisit this long-standing debate
on the static versus dynamic contribution to the flexoelectric
tensor in light of the results presented so far. We shall see that
it bears strong connections to the aforementioned ambiguities in 
the definitions of the continuum fields, and that our formalism
resolves once and for all the existing confusion around this topic.
The coupled dynamics of the polar and acoustic degrees of freedom
is, of course, irrelevant to the study of static structures, such as
the domain walls that we consider in this work. Still, it is interesting
to discuss this topic here, as it provides an additional proof of the
internal consistency of our arguments.

To describe the dynamical evolution of the mechanical and acoustic
degrees of freedom, we need to work out the kinetic energy density in
terms of the mode velocities.
We shall write it as
\black{\begin{equation}
\mathcal{T}({\bf r}) = \frac{1}{2} \sum_{\nu \nu' \alpha\beta} \dot{v}_{\nu\alpha}({\bf r})  M_{\nu\alpha,\nu'\beta} \dot{v}_{\nu'\beta} ({\bf r}),
\end{equation}
where $\alpha\beta$ run over all the Cartesian components of the vector fields indexed
by $\nu\nu'$, and the matrix $M_{\nu\alpha,\nu'\beta}$, of the dimension of a mass density, is 
the normal mode representation of the ``mass operator'' $M$, %atomic masses $m_\kappa$,
\begin{equation}
\begin{split}
M_{\nu\alpha,\nu'\beta} = &
\frac{1}{\Omega} \langle \nu \alpha |M| \nu'\beta \rangle, \\
\langle \kappa \alpha|M|\kappa'\beta\rangle = &
 m_\kappa \delta_{\kappa \kappa'} \delta_{\alpha \beta}.
\end{split}
\end{equation}
(As usual, $\kappa \kappa'$ are sublattice indices; $m_\kappa$ are atomic masses.)}
The off-diagonal kinetic term, coupling the acoustic and optical mode
velocities, has been identified as a \emph{dynamical} contribution to
the bulk flexoelectric tensor by Tagantsev and coworkers~\cite{tagantsev-86,Kvasov-15,Yudin-13}.
%\begin{equation}
% vectors
%over all sublattices.
%projection of the ``atomic distortion 
%fields'' onto the two physically relevant low-energy modes (acoustic
%and polar ``soft mode'').

Based on the arguments of the earlier Sections, it is clear that
the magnitude of such contribution, and even whether it exists at
all, depends on the choice we make for the weights, $w_\kappa$.
The definition given by Tagantsev of the ``static'' flexoelectric
tensor corresponds, within our formalism, to using equal weights 
in the construction of our free energy functional coefficients.
If we do so, the matrix element $M_{uP}$ then reduces to his 
definition of the ``dynamic'' contribution.
If we made a different choice, the partition between the two would 
change arbitrarily -- and yet both the dynamical (phonon
frequencies and dispersions) and static properties (domain
wall energy, equilibrium atomic positions) predicted by our
Lagrangian would be \emph{exactly} the same.
It is interesting to consider the special case where the weights
are set to the physical masses of the atoms divided by the total
mass of the cell, $w_\kappa = m_\kappa/ (\sum_{\kappa'} m_{\kappa'})$.
The orthogonality condition, Eq.~(\ref{ortho}),
immediately leads then to $M_{uP}=0$, i.e., the ``dynamical flexoelectric
effect'' disappears altogether.
Given that the magnitude, and even the very existence, of such 
an effect depends on some arbitrary convention we have made along the way 
in order to map our lattice-dynamical problem onto a continuum Lagrangian 
density, we are forced to conclude that such an effect is not 
measurable.
Still, we find that incorporating a mass cross-term, as suggested
in Ref.~\onlinecite{Yudin-13} is necessary
to guarantee that the physical predictions of the theory are
unaffected by such ambiguities.

%\section{Landau model using the spontaneus polarization}

\begin{table}
\begin{center}
\begin{tabular}{crrrrr}
\hline \hline
& \multicolumn{5}{c}{$\Phi^{(2,xx)}$} \\
\hline
BaTiO$_3$ & $ 0.0920$ & $-0.5561$ & $-0.2022$ & $-0.0383$ & $-0.1299$ \\ 
          & $-0.5561$ & $-0.0191$ & $ 0.0539$ & $-0.0259$ & $-0.0202$ \\ 
          & $-0.2022$ & $ 0.0539$ & $-0.0847$ & $-0.7573$ & $ 0.0142$ \\ 
          & $-0.0383$ & $-0.0258$ & $-0.7573$ & $-0.4640$ & $ 0.1104$ \\ 
          & $-0.1299$ & $-0.0202$ & $ 0.0142$ & $ 0.1104$ & $-0.2174$ \\ 
\hline
BaTiO$_3^{(*)}$ 
          &  0.0915 & -0.5773 & -0.1854 & -0.0319 & -0.1605 \\
          & -0.5773 & -0.0230 &  0.0580 &  0.0177 & -0.0231 \\
          & -0.1854 &  0.0580 & -0.0799 & -0.7699 &  0.0142 \\
          & -0.0319 &  0.0178 & -0.7699 & -0.4921 &  0.1119 \\
          & -0.1605 & -0.0231 &  0.0142 &  0.1119 & -0.2326 \\
\hline
CaTiO$_3$ & $ 0.0082$ & $-0.3563$ & $-0.1513$ & $-0.0596$ & $ 0.1967$ \\ 
          & $-0.3563$ & $-0.0374$ & $ 0.1251$ & $ 0.2363$ & $-0.0124$ \\ 
          & $-0.1513$ & $ 0.1252$ & $-0.1968$ & $-0.9531$ & $ 0.0092$ \\ 
          & $-0.0596$ & $ 0.2363$ & $-0.9531$ & $-0.7121$ & $ 0.1835$ \\ 
          & $ 0.1967$ & $-0.0124$ & $ 0.0092$ & $ 0.1835$ & $-0.0636$ \\ 
\hline
 KNbO$_3$ & $ 0.0195$ & $-0.3345$ & $-0.1154$ & $-0.0194$ & $-0.0530$ \\ 
          & $-0.3345$ & $-0.0671$ & $ 0.1619$ & $ 0.2763$ & $-0.0216$ \\ 
          & $-0.1154$ & $ 0.1619$ & $-0.1782$ & $-0.9074$ & $ 0.0163$ \\ 
          & $-0.0194$ & $ 0.2763$ & $-0.9075$ & $-0.7123$ & $ 0.1230$ \\ 
          & $-0.0530$ & $-0.0216$ & $ 0.0163$ & $ 0.1230$ & $-0.0259$ \\ 
\hline
NaNbO$_3$ & $ 0.0022$ & $-0.2023$ & $-0.0782$ & $-0.0274$ & $ 0.0916$ \\ 
          & $-0.2023$ & $-0.0821$ & $ 0.2043$ & $ 0.3915$ & $-0.0237$ \\ 
          & $-0.0782$ & $ 0.2043$ & $-0.2783$ & $-0.9913$ & $ 0.0250$ \\ 
          & $-0.0274$ & $ 0.3916$ & $-0.9913$ & $-0.8334$ & $ 0.1769$ \\ 
          & $ 0.0916$ & $-0.0237$ & $ 0.0250$ & $ 0.1769$ & $-0.0063$ \\ 
\hline
PbTiO$_3$ & $-0.0529$ & $-0.3944$ & $-0.0856$ & $-0.0003$ & $ 0.1969$ \\ 
          & $-0.3944$ & $-0.0254$ & $ 0.0977$ & $ 0.1092$ & $-0.0240$ \\ 
          & $-0.0856$ & $ 0.0978$ & $-0.1416$ & $-0.7698$ & $-0.0300$ \\ 
          & $-0.0003$ & $ 0.1092$ & $-0.7698$ & $-0.4886$ & $ 0.0226$ \\ 
          & $ 0.1969$ & $-0.0240$ & $-0.0300$ & $ 0.0226$ & $-0.2507$ \\ 
\hline
PbZrO$_3$ & $-0.1344$ & $-0.3815$ & $-0.1047$ & $ 0.0158$ & $ 0.3713$ \\ 
          & $-0.3815$ & $ 0.0260$ & $ 0.0573$ & $ 0.1440$ & $-0.0367$ \\ 
          & $-0.1047$ & $ 0.0573$ & $-0.1189$ & $-0.5866$ & $-0.0134$ \\ 
          & $ 0.0158$ & $ 0.1440$ & $-0.5866$ & $-0.5552$ & $ 0.0543$ \\ 
          & $ 0.3713$ & $-0.0367$ & $-0.0134$ & $ 0.0543$ & $-0.2071$ \\ 
\hline \hline
\end{tabular}
\caption{
Elements of the second-moment interatomic force-constant matrix
(in atomic units), as computed
using DFT calculations in forty-atom $8a_0 \times a_0 \times a_0$ unit cells,
for six perovskite oxides.
}
\label{tab_smifcm}
\end{center}
\end{table}

\section{Supporting numerical data}

%This Appendix contains all the numerical data that we obtained from 
% DFT calculations and it is not explicitely given in the main text.
In Table~\ref{tab_si} we provide the complete list of the calculated model
coefficients for all materials. This is essentially the same data as in 
Table~\ref{tab_coeffs}, only expressed in SI units while setting $P_0$
to the spontaneous ferroelectric polarization of the bulk crystal.
This conversion is useful for two purposes. First, it shows that 
the $A$ and $B$ coefficients are consistent between different weight
choices, provided that the respective distortion vectors, $p_\kappa$,
are related via Eq.~(\ref{ppr}), i.e., they only differ by a sublattice-independent 
constant. 
[This is obviously the case if the electrical polarization is chosen
as a measure of the atomic distortion, but not when the total norm
of the distortion is used; in the latter case there is generally 
an overall scaling factor that originates from Eq.~(\ref{condb}).]
%normalization condition on the polarization 
%distortion vector $p_\kappa$ does not depend on the weights.
%
Second, the coefficients are now expressed in the same units as in
conventional macroscopic theory, allowing for a direct comparison.
In the case of BaTiO$_3$, for example, our calculated polarization
gradient coefficient, $G$, is in good agreement with the phenomenological
value of $G=0.2 \times 10^{-10}$ m$^3$/F reported in Ref.~\onlinecite{Gu-14}. 

The elements of the second-moment interatomic force-constant matrix,
calculated via Eq.~(\ref{sums}), are reported in Table \ref{tab_smifcm} for 
reference.
\bibliography{merged}

\begin{thebibliography}{53}
\bibitem [{lau()}]{lautrec}%
  \BibitemOpen
  \href@noop {} {}\bibinfo {note} {{\sc Lautrec} is a massively parallel
  ``in-house" electronic-structure code currently maintained by Massimiliano
  Stengel}\BibitemShut {NoStop}%
\end{thebibliography}

\end{document}